\documentclass[twocolumn,showpacs,pra,eqsecnum,citeautoscript,amsmath,amssymb,floatfix,superscriptaddress]{revtex4-1}
\usepackage{bm,xcolor,amsmath,amssymb,mathrsfs,latexsym,graphicx,psfrag,float}
\usepackage{dcolumn}
\usepackage{hyperref}
\usepackage{leftidx}

\usepackage{tabularx}

\usepackage[english]{babel}
\usepackage[utf8]{inputenc}
\usepackage{bbold}
\usepackage{dsfont}

\usepackage{bm}        
\usepackage{filecontents}
\usepackage{lineno}
\newcommand{\cre}[2]{ #1^{\dagger}_{#2}}
\newcommand{\cref}[2]{ #1^{\ast}_{#2}}
\newcommand{\ann}[2]{ #1^{\phantom{\dagger}}_{#2}}

\usepackage{bbm}

\usepackage{xcolor}

\newcommand{\sub}[1]{_{\text{#1}}}

\DeclareMathOperator{\sign}{sign} 

\begin{document}

\title{Thermalization Dynamics of Two Correlated Bosonic Quantum Wires After a Split}
\author{Sebastian Huber}
\affiliation{Physics Department, Arnold Sommerfeld Center for Theoretical Physics and Center for NanoScience, Ludwig-Maximilians-University Munich, 80333 Munich, Germany}
\author{Michael Buchhold}
\affiliation{Institute of Theoretical Physics, University Cologne, 50937 Cologne, Germany}
\affiliation{Department of Physics and Institute for Quantum Information and Matter, California Institute of Technology, Pasadena, CA91125, USA}
\author{J\"org Schmiedmayer}
\affiliation{Vienna Center for Quantum Science and Technology, Atominstitut, TU Wien, Stadionallee 2, 1020 Vienna, Austria}
\author{Sebastian Diehl}
\affiliation{Institute of Theoretical Physics, University Cologne, 50937 Cologne, Germany}

\begin{abstract}
Coherently splitting a one-dimensional Bose gas provides an attractive, experimentally established platform to investigate many-body quantum dynamics. At short enough times, the dynamics is dominated by the dephasing of single quasi-particles, and well described by the relaxation towards a generalized Gibbs ensemble corresponding to the free Luttinger theory. At later times on the other hand, the approach to a thermal Gibbs ensemble is expected for a generic, interacting quantum system. Here, we go one step beyond the quadratic Luttinger theory and include the leading phonon-phonon interactions. By applying kinetic theory and non-equilibrium Dyson-Schwinger equations, we analyze the full relaxation dynamics beyond dephasing and determine the asymptotic thermalization process in the two-wire system for a symmetric splitting protocol. The major observables are the different phonon occupation functions and the experimentally accessible coherence factor, as well as the phase correlations between the two wires. We demonstrate that, depending on the splitting protocol, the presence of phonon collisions can have significant influence on the asymptotic evolution of these observables, which makes the corresponding thermalization dynamics experimentally accessible.  
\end{abstract}

\maketitle

\section{Introduction}\label{A1}
In recent years, the field of ultracold atoms has produced tremendous progress in the experimental realization and observation of many-body quantum dynamics out of equilibrium \cite{kinoshita2006quantum, gring2012relaxation, langen2013local, hofferberth2007non, trotzky2012probing, meinert2013quantum, geiger2014local}. Experiments have demonstrated that the purity of cold atom systems in combination with their large degree of controllability allow one to precisely probe the fundamental processes in quantum many-body systems. This is exploited by a growing number of experiments, which study the relaxation dynamics of interacting quantum systems in a controlled environment, thereby fostering a more detailed understanding of paradigmatic single- and many-body effects, such as dephasing \cite{andrews1997observation,kinoshita2006quantum,bistritzer2007intrinsic,whitlock2003relative} and thermalization \cite{hofferberth2007non,schachenmayer2014spontaneous,strohmaier2010observation,rigol2008thermalization,sieberer2016keldysh,foini2015nonequilibrium}.

One-dimensional quantum systems play a special role in this context since the dynamics, especially that of collisions, is constrained to a reduced phase space. For many systems this leads to the emergence of an extensive number of quasi-conserved charges, i.e. an extensive number of physical observables  which are protected from fast relaxation by approximate conservation laws. The violation of these conservation laws becomes only visible for large time scales. Consequently, a large number of experiments with one-dimensional quantum gases has been described very well in terms of integrable models, for which the conservation laws become exact \cite{geiger2014local, gring2012relaxation, langen2015experimental, haller2010pinning}. One general feature of these integrable models is the absence of thermalization, expressed by the fact that, due to the large number of conservation laws, they never relax towards a thermal Gibbs ensemble \cite{jaynes1957information, deutsch1991quantum, srednicki1994chaos, srednicki1999approach, rigol2008thermalization, rigol2007relaxation, geiger2014local, gring2012relaxation, langen2015experimental,berges2004prethermalization}, but keep the information on their initial state for all times and evolves towards a so-called generalized Gibbs ensemble (GGE).

Realistic systems, however, are expected to behave generically, and to asymptotically approach a thermal state, although very slowly in one dimension \cite{kollar2011generalized, buchhold2016prethermalization}. The aim of this work is to study the full relaxation dynamics process, including thermalization, for a gas of ultracold bosons after initialization in an out-of-equilibrium state. The specific setup consists of an initial, one-dimensional Bose-Einstein condensate (BEC), which is suddenly split into two halves along its longitudinal axis, as it is performed experimentally in Vienna \cite{kitagawa2011dynamics,geiger2014local, gring2012relaxation, langen2015experimental}. The  system, now consisting of two initially strongly correlated condensates, evolves in time and relaxes towards a stationary state. 

On short enough time scales, most of the relaxation dynamics has been described successfully in terms of the integrable Luttinger theory. It predicts the light-cone shaped dephasing of anomalous phonon modes, which leads to the transient relaxation of the system towards a GGE \citep{jaynes1957information,rigol2007relaxation,polkovnikov2011colloquium,kollar2011generalized}. The GGE state is determined by the initial expectation value of the conserved charges of the integrable Luttinger theory. This description, however, neglects the effect of phonon-phonon collisions and the associated decay of the inter-wire correlations, as well as the energy redistribution between the modes within each individual wire. The processes, while very slow on experimental time scales, lead to the breakdown of the GGE on larger time scales, and the relaxation of the system towards a thermal Gibbs state. 

The goal of this paper is to obtain a complete and quantitative picture of the thermalization dynamics of one-dimensional Bose gases, including the regime where non-linear effects are important. In particular, we assess the observability of the associated physical effects in experiments. To this end, we study the time evolution of the two-wire system after the splitting in terms of a non-linear Luttinger theory, which incorporates the leading order phonon scattering mechanisms for a gas of one-dimensional bosons. We focus on a regime with a large Luttinger parameter $K\gg1$, i.e. weakly interacting bosons, for which diagrammatic methods are applicable \cite{buchhold2015nonequilibrium, buchhold2016prethermalization, buchhold2015kinetic, andreev1980hydrodynamics, burkov2007decoherence, punk2006collective}. The BEC is longitudinally trapped in a box potential of length $L$, which can be realized experimentally with high accuracy \cite{Rauer2017} and which allows us to approach the dynamics within a Luttinger framework with a physical infrared momentum cutoff $k_{\text{IR}}=L^{-1}$. All analytical results will be discussed for $k_{\text{IR}}=0$, while the numerical simulations are carried out for $k_{\text{IR}}=L^{-1}$. We determine the time-evolution of temporally local correlators of the system, as these are the quantities accessible in current experiments.  

{\it Summary of main results.} -- Our approach is based on a combination of Dyson-Schwinger and kinetic equations for the weakly interacting Bose gas. In this way, we are able to quantitatively capture all stages of the thermalization process, including the hydrodynamic relaxation enforced by global conservation laws on asymptotically large times \cite{buchhold2016prethermalization}.

At short times, dephasing of non-interacting phonon modes leads to a fast relaxation towards a generalized Gibbs ensemble. The physics can be understood on the single phonon level, i.e. technically in the framework of the quadratic Luttinger theory.  For intermediate and large times $t>t_{\text{coll}}$, set by the typical phonon collision time $t_{\text{coll}}$, interactions become important and relaxation dynamics enters the realm of many-body physics. Collisions lead to energy redistribution between the phonons and to local equilibration, while on the other hand conservation laws do not yet play a dominant role. On the asymptotic timescales, however, conservation laws constrain the effects of many collisions and the dynamics enters a many-body hydrodynamic regime. This leads to algebraically slow relaxation of observables \cite{buchhold2016prethermalization}.

In addition to a quantitative description of the different dynamical regimes, we obtain as well a time resolved estimate for the crossover scale $x_t^{(c)}\sim t^{2/3}$ between the single-particle and the many-particle regime. It grows in time with the typical Andreev exponent $1/\alpha=2/3$ \cite{buchhold2015kinetic, andreev1980hydrodynamics, burkov2007decoherence, punk2006collective}, confirming the picture that thermalization proceeds from high to low momentum modes, with crossover momentum $q^{(c)} \sim 1/x^{(c)}$. Collisions also lead to a stronger decay of coherence, which modifies the decay of the relative phase correlation function $\log C_r \sim -t^{4/3}$ to faster than linear in time on large distances.

Our formulation allows us to address direct experimental observables, such as the relative phase correlation function and the coherence factor. We find that a careful choice of observables is necessary to quantitatively certify thermalization in the experiment. In more detail, our results indicate that for the standard splitting procedure both the relative phase correlation function as well as the coherence factor are only weakly modified by phonon collisions beyond the quadratic Luttinger model. Thermalization, while definitely present, is therefore hard to observe in common observables. This is caused by the form of the initial phonon distribution function, which follows a thermal Rayleigh-Jeans divergence $n_{p,t}\sim T_r/|p|$, with typical temperature $T_r$, at low momenta. While this temperature increases through the thermalization process towards an absolute temperature $T_a$, the form of the distribution remains untouched and thus no qualitative change in observables can be measured. 

In order to overcome the difficulty in resolving thermalization, we propose to initialize the condensate in a modified far-from equilibrium state, for which the phonon densities are different from a thermal distribution. Inspired by the recent realization of one such non-thermal state \cite{langen2015experimental}, we investigate initial states for which the phonon distribution oscillates on momentum scales $\xi^{-1}$. These oscillations remain preserved under dephasing but are washed out by the redistribution of energy in the thermalization process, leading to a strongly increased signal of thermalization at distances $x=2\xi$, which could be detected in experiments and which would lead to a clear observation of $t_{\text{coll}}$.

This paper is organized as follows. We start by introducing the interacting Luttinger model and discussing the relevant physical observables in Sec.~\ref{A2}. In Sec.~\ref{A4}, we derive the kinetic equation approach for the correlated two-wire system. This section can be skipped by a reader more interested in the physics results. The formalism is then applied in Sec.~\ref{A3} to study the relaxation dynamics of an initial state which has nearly thermal form, corresponding to Refs.~\cite{geiger2014local, gring2012relaxation}. A discussion of the various stages of the relaxation process is provided in subsection~\ref{A5}. In Sec.~\ref{A8}, a class of initial states is considered, which differ more strongly from a thermal state, and a representative of which has been implemented in recent experiments~\cite{langen2015experimental}. We conclude in Sec.~\ref{Conc}.

\section{Model}\label{A2}
The present work analyzes the longitudinal splitting of a one-dimensional Bose-Einstein condensate at time $t=0$, and the subsequent time evolution of the two resulting, identical and initially correlated BECs \cite{kitagawa2011dynamics,gring2012relaxation,geiger2014local}. Each individual wire represents a one-dimensional gas of interacting bosons, whose long wavelength dynamics will be described in the following  in terms of an interacting Luttinger liquid theory. For a sufficiently strong separation of the two wires, all inter-wire couplings in the Hamiltonian vanish and the only correlations between the two gases result from the correlations in the initial state for $t<0$. 

\subsection{Hamiltonian}
Labeling the two subsystems with bosonic creation and annihilation operators $\cre{b}{\alpha,x}, \ann{b}{\alpha,x}$, $\alpha\in\{1,2\}$, where $\ann{b}{\alpha,x}$ annihilates a boson at coordinate $x$ in wire $\alpha$,
the Hamiltonian of the joint system for times $t>0$ is
\begin{align}
H = H_{1,\text{IB}} \otimes  \mathds{1}_{2} + \mathds{1}_{1} \otimes H_{2,\text{IB}}.\label{E2}
\end{align}
Here, $H_{\alpha,\text{IB}}$ describes a generic system of short-range interacting bosons with mass $m$,
\begin{align}
H_{\alpha,\text{IB}}=\int_x\cre{b}{\alpha,x}\left[V(x)-\frac{\hbar^2}{2m}\nabla^2\right]\ann{b}{\alpha,x}\nonumber\\
+\int_{x,x'}\cre{b}{\alpha,x}\ann{b}{\alpha,x}u(x-x')\cre{b}{\alpha,x'}\ann{b}{\alpha,x'},\label{Ham}
\end{align}
with a two-body interaction potential $u$, which is short ranged in space and an external potential $V$, which we consider flat enough to justify a local density approximation (LDA).

The creation and annihilation operators are conveniently expressed in terms of conjugate phase and amplitude fields, $b_{\alpha,x}= \sqrt{\rho_{\alpha,x}}e^{i\theta_{\alpha,x}}$, with $\rho_{\alpha,x}=\rho_{0}-\frac{1}{\pi}\nabla\phi_{\alpha,x}$, where $\rho_0$ is the average particle density in both wires \citep{giamarchi2004quantum,haldane1981effective}. Phase and amplitude are conjugate variables
\begin{align}
[\phi_{\alpha,x},\nabla\theta_{\beta,x'}]=-i\delta(x-x')\delta_{\alpha,\beta}.\label{E1}
\end{align}

In phase-amplitude representation, the leading order terms of the Hamiltonian \eqref{Ham} correspond to an interacting Luttinger liquid (ILL) \cite{andreev1980hydrodynamics,buchhold2015nonequilibrium,buchhold2015kinetic} and reads
\begin{align}
H_{\alpha,\text{ILL}}=\tfrac{\hbar}{2\pi}\int_x\big[\nu K\left(\partial_x\theta_{\alpha,x}\right)^2+\tfrac{\nu}{K}\left(\partial_x\phi_{\alpha,x}\right)^2\nonumber\\+\kappa\left(\partial_x\phi_{\alpha,x}\right)\left(\partial_x\theta_{\alpha,x}\right)^2
\big].\label{Ham2}
\end{align}
The first line of this equation is quadratic in the fields and represents the non-interacting Luttinger liquid, which describes non-interacting elementary excitations, i.e. phonons, propagating with sound velocity $\nu=\sqrt{\tfrac{\rho_0 u(0)}{m}}$ and Luttinger parameter $K=\tfrac{\pi\hbar}{2}\sqrt{\frac{\rho_0}{u(0)m}}$. This quadratic part is integrable and governs the  transient dynamics after the splitting. It leads to the evolution of the system towards a quasi-stationary state, a so-called generalized ensemble (GGE) \citep{jaynes1957information,rigol2007relaxation,polkovnikov2011colloquium,kollar2011generalized}. 

The second line of Eq.~\eqref{Ham2} instead is a non-linear coupling, which represents the leading order deviation from an integrable model and makes the ILL generic, i.e. it enables true thermalization on large and asymptotic time scales \cite{andreev1980hydrodynamics,buchhold2015nonequilibrium,buchhold2015kinetic}. The vertex $\kappa$ describes the deviation from a perfectly linear dispersion and can be estimated to be $\kappa=\hbar/m$, where $m$ is the mass of an atom. The major contribution of this work is to include the effect of non-zero $\kappa$, and to study the asymptotic thermalization after two different, experimentally relevant splitting procedures. This extends the understanding of the relaxation dynamics after the splitting to time scales beyond prethermalization.

The Hamiltonian of the entire system after the splitting is again the sum of the two individual wires 
\begin{align}
H = H_{1,\text{ILL}} \otimes  \mathds{1}_{2} + \mathds{1}_{1} \otimes H_{2,\text{ILL}}.\label{Ham3}
\end{align}
Since the considered splitting protocol is symmetric with respect to the wires, both will be identical in geometry and average particle density after the splitting. Thus, they are described by the same set of Luttinger parameters $\{\nu, K, \kappa\}$. 

In order to diagonalize the quadratic part of the Hamiltonian $H_{\alpha,\text{ILL}}$, one applies the canonical Bogoliubov transformation \cite{giamarchi2004quantum,haldane1981effective} to the phonon operators $\cre{a}{\alpha,x}, \ann{a}{\alpha,x}$,
\begin{align}
\theta_{\alpha,x}&=\Theta_{\alpha}+i\hspace{-0.1cm}\int_p' \hspace{-0.1cm}\sqrt{\frac{\pi}{2 K |p|}}e^{-ipx-\frac{|p|}{\Lambda}}(a^{\dagger}_{\alpha,p}-a_{\alpha,-p}),\label{E6}\\
\phi_{\alpha,x}&=\Phi_{\alpha}-i\hspace{-0.1cm}\int_p'\hspace{-0.1cm}\sqrt{\frac{\pi K}{2|p|}} \sign (p) e^{-ipx-\frac{|p|}{\Lambda}}(a^{\dagger}_{\alpha,p}+a_{\alpha,-p}),\label{E7}
\end{align}
where $\Lambda=\sqrt{4mu(0)\rho_0}/\hbar$ is the short-distance cutoff, $\ann{a}{\alpha,p}=\int_x e^{ipx}\ann{a}{\alpha,x}$ and the integral $\int'_p$ runs over all momenta except $p=0$. The global, zero momentum modes $\Theta_\alpha, \Phi_\alpha$ contribute only to the global dephasing of the wires \cite{Rauer2017}, which is not considered here, and are therefore neglected in the following.

In this basis, the quadratic part of the Hamiltonian (first line of Eq.~\eqref{Ham2}) is diagonal and describes phonons with a linear dispersion $\epsilon_p=\nu|p|$. The non-linear part of the Hamiltonian describes cubic phonon scattering processes. It can be decomposed into a resonant part, which conserves the phonon energies, and an off-resonant part, which does not conserve the phonon energy during a single scattering process. The latter processes induce only virtual transitions between the many-body states and represent irrelevant contributions for the forward time evolution, in contrast to the resonant ones \cite{kamenev2011field}. The common procedure is thus to neglect the off-resonant scattering processes in a rotating wave approximation for the kinetics at large times, and only consider resonant processes. In the resonant approximation, the Hamiltonian is
\begin{align}
H_{\alpha,\text{ILL}}=&\int_p \nu|p|\cre{a}{\alpha,p}\ann{a}{\alpha,p}\nonumber\\
&+v\int''_{p,q}\sqrt{|pq(q+p)|}\left(\cre{a}{\alpha,p+q}\ann{a}{\alpha,p}\ann{a}{\alpha,q}+\text{H.c.}\right),\label{Ham8}
\end{align} 
with $v=\kappa\sqrt{\tfrac{9\pi}{2K}}$.
The integral $\int''_{q,p}$ only considers resonant processes, i.e. those processes with $|p+q|=|p|+|q|$ (for more details on the resonant approximation, see \cite{kamenev2011field, buchhold2015nonequilibrium, buchhold2015kinetic}). From now on, we set $\hbar=1$ within the analytic expressions.

The Hamiltonian, Eqs.~\eqref{Ham3} and \eqref{Ham8}, describes the time evolution of the two wires after they have been split initially at $t=0$. Each wire thereby represents an independent, interacting Luttinger liquid and the Hamiltonian generates no additional inter-wire correlations. The latter are, however, generated by the splitting procedure itself and the analysis of the inter-wire and intra-wire correlations' time evolution under \eqref{Ham3} is the purpose of this article.

\subsection{Phonon correlation functions}

In the following sections, the dynamics of the two wires is analyzed in terms of the single-particle phonon Green's functions, which in turn determine a set of relevant physical observables.  These are for instance the so-called phase correlation function $C(x,t)=\langle e^{i(\theta_{1,x,t}-\theta_{2,x,t})}e^{-i(\theta_{1,0,t}-\theta_{2,0,t})}\rangle$ between the two wires and the coherence factor $\Psi_c(t)$, which we will introduce in this section. The basic notation and relations for the non-equilibrium Green's functions will be discussed below.

In the Heisenberg picture, with operators $\ann{a}{\alpha,p,t}=e^{iHt}\ann{a}{\alpha,p}e^{-iHt}$, the retarded Nambu response function is defined as
\begin{align}
iG_{\alpha,\beta,p,t,t'}^{R}\hspace{-0.1cm}=\hspace{-0.1cm}
\Theta(t-t')\hspace{-0.1cm}\begin{pmatrix}
  \langle[\ann{a}{\alpha,p,t},\cre{a}{\beta,p,t'}]\rangle& \langle[\ann{a}{\alpha,p,t},\ann{a}{\beta,-p,t'}]\rangle \\ \langle[\cre{a}{\alpha,-p,t},\cre{a}{\beta,p,t'}]\rangle& \langle[\cre{a}{\alpha,p,t},\ann{a}{\beta,p,t'}]\rangle
 \end{pmatrix},\label{Cor1}
\end{align}
while the Nambu correlation function (or Keldysh Green's function) is defined as
\begin{align}
iG_{\alpha,\beta,p,t,t'}^{K}=
\begin{pmatrix}
  \langle\{\ann{a}{\alpha,p,t},\cre{a}{\beta,p,t'}\}\rangle& \langle\{\ann{a}{\alpha,p,t},\ann{a}{\beta,p,t'}\}\rangle \\ \langle\{\cre{a}{\alpha,p,t},\cre{a}{\beta,p,t'}\}\rangle& \langle\{\cre{a}{\alpha,p,t},\ann{a}{\beta,p,t'}\}\rangle
 \end{pmatrix}.\label{Cor2}
\end{align}
Here, $[\cdot,\cdot]$ labels the commutator and $\{\cdot,\cdot\}$ the anti-commutator.

The retarded Green's function can immediately be simplified by noticing, that the Hamiltonian \eqref{Ham3} does not introduce any inter-wire coupling and therefore states with $\alpha\neq\beta$ in Eq.~\eqref{Cor1} commute at all times. Furthermore, within a single wire, $H_{\alpha,\text{ILL}}$ in \eqref{Ham8} does not couple states with opposite momenta $\ann{a}{\alpha,q,t}, \ann{a}{\alpha,-q,t'}$. Consequently, the off-diagonal elements of $G^R$ vanish exactly for all times $t, t'$. The advanced Green's function is the hermitian conjugate of the retarded Green's function, $G^A=\left(G^R\right)^{\dagger}$.

A special emphasis will be put on the equal time Green's functions, relevant for current experiments, which determine the time evolution of static, i.e. frequency independent observables. For the retarded and advanced Green's functions, one finds
\begin{align}
\left(G^R_{\alpha,\beta,q,t,t}-G^A_{\alpha,\beta,q,t,t}\right)&=-i\delta_{\alpha,\beta}\sigma_z,\\ \left(G^R_{\alpha,\beta,q,t,t}+G^A_{\alpha,\beta,q,t,t}\right)&=0
\end{align}
for all times, while 
\begin{align}
iG_{\alpha,\beta,p,t,t}^{K}=
\begin{pmatrix}
  2n_{\alpha,\beta,p,t}+\delta_{\alpha,\beta}& 2m_{\alpha,\beta,p,t}e^{2i\nu|p|t} \\ 2m_{\alpha,\beta,p,t}e^{-2i\nu|p|t}& 2n_{\beta,\alpha,p,t}+\delta_{\alpha,\beta}\end{pmatrix}.\label{Cor2}
\end{align}
It describes the time evolution of intra- ($\alpha=\beta$) and inter-wire ($\alpha\neq\beta$) phonon densities, both diagonal $n_{\alpha,\beta,p,t}=\langle\cre{a}{\alpha,p,t}\ann{a}{\beta,p,t}\rangle$ and off-diagonal $m_{\alpha,\beta,p,t}=|\langle\ann{a}{\alpha,p,t}\ann{a}{\beta,p,t}\rangle|$. Here we made the time-dependent complex phase of the off-diagonal modes explicit such that $m_{\alpha,p,t}$ is real and positive for all times.

Both wires are exactly identical and therefore the system is symmetric under the exchange of the wire index, this yields 
\begin{align}
G^{R/A/K}_{11,p,t,t'}=G^{R/A/K}_{22,p,t,t'}\  \text{ and }\ G^{R/A/K}_{12,p,t,t'}=G^{R/A/K}_{21,p,t,t'}.\label{Rel1}
\end{align}
The same holds for the phonon densities in the wire basis
\begin{align}
n_{11,p,t}=\langle a^{\dagger}_{1,p,t}a_{1,p,t}^{\phantom{\dagger}}\rangle=\langle a^{\dagger}_{2,p,t}a^{\phantom{\dagger}}_{2,p,t}\rangle=n_{22,p,t},\label{Partnum1}\\
n_{12,p,t}=\langle a^{\dagger}_{1,p,t}a_{2,p,t}^{\phantom{\dagger}}\rangle=\langle a^{\dagger}_{2,p,t}a^{\phantom{\dagger}}_{1,p,t}\rangle=n_{21,p,t},\\
m_{12,p,t}=|\langle a^{\phantom{\dagger}}_{1,p,t}a_{2,p,t}^{\phantom{\dagger}}\rangle|=|\langle a^{\phantom{\dagger}}_{2,p,t}a^{\phantom{\dagger}}_{1,p,t}\rangle|=m_{21,p,t}.
\end{align}

\subsection{Relative and absolute mode and experimental observables}
Due to the symmetry of the system under exchange of the wires, i.e. $1\leftrightarrow 2$, the Green's functions in the wire basis $G^{R/A/K}_{\alpha,\beta}$ show some redundancy. It can be removed by switching from the basis of individual wires to a basis of relative and absolute degrees of freedom for the two wires. These latter degrees of freedom give a more compact description and are directly related to the relevant experimental observables such as the relative phase correlation function and the coherence factor of the bosons. 

The relative ($ \ann{a}{r,p,t}$) and absolute ($ \ann{a}{a,p,t}$) degrees of freedom are introduced by the unitary transformation of the Heisenberg operators
\begin{align}
 \begin{pmatrix}
  \ann{a}{a,p,t} \\ 
  \ann{a}{r,p,t}
 \end{pmatrix}
 =\frac{1}{\sqrt{2}}
\begin{pmatrix}
  1 & 1 \\ 
  1 & -1
 \end{pmatrix}
 \begin{pmatrix}
  \ann{a}{1,p,t} \\ 
\ann{a}{2,p,t}
 \end{pmatrix} \label{E12}.
\end{align}\\
Using Eq. \eqref{Rel1}, one can directly see that correlations between the absolute and relative modes vanish, and that only two independent Nambu Green's functions are required, namely the absolute and the relative Green's functions
\begin{align}
G^{\lambda}_{a,p,t,t'}&=G^{\lambda}_{11,p,t,t'}+G^{\lambda}_{12,p,t,t'},\\
G^{\lambda}_{r,p,t,t'}&=G^{\lambda}_{11,p,t,t'}-G^{\lambda}_{12,p,t,t'},
\end{align}
with $\lambda=R,A,K$. We note that this does not require any specific form of the interactions but only the fact that the Hamiltonian as well as the initial conditions are invariant under exchange of the wire indexes.

The dynamics of $G^{\lambda}_{r/a,p,t,t'}$ is uniquely determined by $G^{\lambda}_{11,p,t,t'}$ and $G^{\lambda}_{12,p,t,t'}$. It is equivalent to work in the relative and absolute basis or in the basis of individual wires. The Hamiltonian, however, and the corresponding time evolution takes a simpler form in the individual wire basis. On the other hand, experimental signatures and the initial state of the system are determined in the relative-absolute basis. In the following, the most convenient of both representations is used, depending on the individual computation.

The major experimental observable for the present system is the relative phase between the two wires, from which the relative coherence between the wires is extracted. This observable can be measured with high precision in matter-wave interferometry measurements \citep{polkovnikov2006interference,bistritzer2007intrinsic,burkov2007decoherence}. This is done by releasing the two condensates from the wires, and an letting them expand freely in space. When the two condensates start to overlap, they form a characteristic interference pattern. The integrated interference pattern is described by the coherence factor of the two condensates
\begin{align}
\Psi_c(t)=L^{-1}\int_x\langle \cre{b}{1,x,t}\ann{b}{2,x,t}\rangle =\frac{\rho_0}{L}\int_x\langle e^{i(\theta_{r,x,t})}\rangle=\rho_0e^{-g(t)},\label{phid}
\end{align}
where the integral runs over the entire length $L$ of the wire and $\theta_{r,x,t}=\theta_{1,x,t}-\theta_{2,x,t}$ is the relative phase of the two condensates. The subleading effect of density fluctuations has been neglected in the definition of $\Psi_c(t)$.

According to the linked cluster theorem, the coherence factor up to fourth order vertex corrections \cite{mahan2013many,buchhold2016prethermalization} is 
\begin{align}
g(t)&=\int_p\frac{\pi e^{-\frac{|p|}{\Lambda}}}{8K|p|}\text{Tr}\left[iG^K_{r,p,t}(\mathds{1}+\sigma^x)\right]\label{phid2}\\
&=\int_p\frac{\pi e^{-\frac{|p|}{\Lambda}}}{8K|p|}\left(2n_{r,p,t}+1-2\cos(2\nu|p|t)m_{r,p,t}\right).\nonumber
\end{align}
It is fully determined by the time evolution of the relative phonon densities $n_{r,p,t}, m_{r,p,t}$. For the present analysis, we approximate the system as being  spatially homogenous. This approximation applies very well for current experiments, which are governed by the dynamics in the longitudinal center of the wires \cite{kitagawa2011dynamics,geiger2014local,gring2012relaxation}.

A further tool to get experimental access to the correlations between the two wires is to measure the relative phase between the two wires for different positions in space. The observed interference pattern of this measurement is proportional to the phase correlation function
\begin{align}
C(x,t) =  e^{-\frac{1}{2} \Delta \theta_{r,x,t}},\label{E47}
\end{align}
where $x=x_1 - x_2$ and $\Delta \theta_{r,x,t}=\langle [\theta_{r,x,t}-\theta_{r,0,t}]^{2}\rangle$. The relative phase fluctuation $\Delta\theta_{r,x,t}$ is determined along the lines of Eq.~\eqref{phid},
\begin{align}
\Delta \theta_{r,x,t}=\int_p\Big[&\frac{\pi e^{-\frac{|p|}{\Lambda}}}{4Kp}(1-\cos(px))\nonumber\\
&\left(2n_{r,p,t}+1-2\cos(2\nu|p|t)m_{r,p,t}\right)\Big].\label{E48}
\end{align}
It reveals the time-dependent spatial spread (or decay) of correlations and has served as the major experimental observable to monitor the time-dependent relaxation of the two-wire system. In order to shorten notation in the following, we will refer to $\Delta\theta_{r,x,t}$ simply as the relative phase.

In order to make theoretical predictions on the dynamics of the phase correlation function and the coherence factor of the system after the split, we thus need to determine the time evolution of the relative phonon densities $n_{r,p,t}, m_{r,p,t}$. In the major body of previous works, this has been performed in terms of the quadratic Luttinger theory, which neglects collisions between phonons, as described by the cubic part of Eq.~\eqref{Ham8}. This is well justified for short and intermediate times, which are smaller that the typical, inverse collision rate, and describes the so-called prethermalization dynamics.
The present work goes beyond the former, non-interacting approaches and takes into account the effect of phonon-phonon scattering, which leads to an energy redistribution between the phonon modes and enables thermalization on large time scales. The time-evolution of the phonon densities is determined via a kinetic equation approach, which is set up in the following section.

\section{Kinetic equation for the phonon densities}\label{A4}
The time evolution of the phonon densities in the present system is determined by a modified version of the common kinetic equation approach \cite{kamenev2011field,buchhold2015kinetic}. It has been established in previous works in order to deal with the resonant phonon interactions, for which perturbative approaches display fatal on-shell divergencies. The kinetic equation approach here is based on Dyson-Schwinger equations \cite{amit2005field,peskin1995introduction}, resulting in regular expressions for the time evolution of the phonon densities. The Dyson-Schwinger equations will be truncated at the level of the cubic vertex (to be precise, including the cubic vertex correction) and solved self-consistently. This approach has been outlined in detail in \cite{buchhold2015nonequilibrium,buchhold2015kinetic} and will be briefly discussed below, including the extension to the bosonic two-wire system.

The kinetic equation is expressed in temporal Wigner coordinates, which describe the time evolution of any two-time function, depending on time variables $t_1, t_2$, as a function of the average forward time $t=\frac{t_1+t_2}{2}$ and the relative time $\delta_t=t_1-t_2$. Fourier transforming the relative time coordinate introduces the frequency $\omega$, i.e. $G_{t,\omega}=\int_{\delta_t}e^{i\omega\delta_t}G_{t,\delta_t}$ for an arbitrary Green's function. Frequency integration thus produces equal-time functions in Wigner representation.
The Dyson equation \cite{mahan2013many} in the individual wire basis and in Wigner coordinates reads
\begin{align}
G_{\alpha\beta,p,\omega,t}^{-1}=\begin{pmatrix}
  0 & P^{A}_{\alpha\beta,p,\omega,t}-\Sigma^{A}_{\alpha\beta,p,\omega,t} \\ 
P^{R}_{\alpha\beta,p,\omega,t}-\Sigma^{R}_{\alpha\beta,p,\omega,t} & -\Sigma^{K}_{\alpha\beta,p,\omega,t}
 \end{pmatrix},\label{E20}
\end{align}
with the self-energies $\Sigma^{R,A,K}_{\alpha\beta,p,\omega,t}$, and the bare propagators of the non-interacting system $P^{R,A}_{\alpha\beta,p,\omega,t}=\left(\sigma_{z}(\omega \pm i\eta)-\nu|p|\mathds{1}\right)\delta_{\alpha,\beta}$.
The retarded and advanced Green's functions are modified in the presence of interactions due to non-zero self-energies $\Sigma^{R,A}_{\alpha\beta,p,\omega,t}$. As pointed out above, however, the Hamiltonian neither mixes different wires $\alpha\neq\beta$ nor opposite momenta $p, -p$ and thus the retarded and advanced self-energies are diagonal in the wire index and in Nambu space.

Considering a spectral function that has support only in the close vicinity of the dispersion, the relevant self-energy contributions for the kinetic equation are the ones, which are on-shell, i.e. which fulfill $\omega=\nu|p|$ \cite{kamenev2011field}. Here, these contributions are purely imaginary and the on-shell self-energy can be parametrized as
 \cite{buchhold2015kinetic} (see also appendix \ref{AS1}: Eq.~\eqref{ES15} with $\Phi_{\alpha\beta,p,\omega,t}^{R}=0$)
\begin{align}
\Sigma^{R}_{\alpha\beta,p,\omega=\nu|p|,t}=-i\delta_{\alpha\beta}
\begin{pmatrix}
\sigma^{R}_{p,t} & 0\\
0 &\sigma^{A}_{p,t}\end{pmatrix}.\label{E22}
\end{align}
Here, $\sigma^{R,A}_{p,t}$ are positive functions of momentum $p$ and forward time $t$. For $\sigma^{R,A}_{p,t}\ll \nu|p|$, the self-energy is interpreted as the inverse lifetime $\tau_{p,t}^{-1}=\sigma^R_{p,t}$ of the phonon modes, which remain well defined quasi-particles in this case.

The on-shell Keldysh self-energy is, by definition, anti-hermitian and, due to the symmetry of the system with respect to exchange of the wires, conveniently parametrized as
\begin{align}
\Sigma^{K}_{\alpha\beta,p,\omega=\nu|p|,t}&=-i2\begin{pmatrix}
\sigma_{\alpha\beta,p,t}^{K} & \Phi_{\alpha\beta,p,t}^{K}\\
\Phi_{\alpha\beta,p,t}^{K}& \sigma_{\alpha\beta,p,t}^{K}\end{pmatrix}.\label{E23}
\end{align}
In contrast to the retarded self-energy, both the off-diagonal elements of the Keldysh self-energy in Nambu space, as well as in the wire index, are generally non-vanishing. This is a consequence of the initial splitting protocol, which generates correlations between the wires but also occupies anomalous phonon modes. We have, however, again factored out the complex phase of the off-diagonal modes, as in Eq.~\eqref{Cor2}. As a consequence, $\Phi^K$ is purely real for all times.

The kinetic equation for the phonon occupations is determined by parametrizing the anti-hermitian Keldysh Green's function in terms of the retarded, advanced Green's functions and the hermitian occupation function $F$, i.e.
\begin{align}
G^{K}_{p,\omega,t}=\left(G^R\circ \Sigma_z F-F\Sigma_z\circ G^A\right)_{p,\omega,t},\label{Par}
\end{align}
where $\circ$ denotes the convolution with respect to times and frequencies and matrix multiplication with respect to momentum, Nambu and wire index. The matrix $\Sigma_z=\sigma_z\otimes\mathds{1}$ ensures the symplectic structure of bosonic Nambu space. Inserting Eq.~\eqref{Par} into \eqref{E20}, one finds
\begin{align}
\left(P^{R} \circ \Sigma_{z}F-F\Sigma_{z} \circ P^{A}\right)_{p,\omega,t}=I^{\text{coll}}_{p,\omega,t},\label{E24}
\end{align}
with the collision integral
\begin{align}
I^{\text{coll}}_{p,\omega,t}=\Sigma^{K}_{p,\omega,t}-\left(\Sigma^{R} \circ \Sigma_{z}F-F\Sigma_{z} \circ \Sigma^{A}\right)_{p,\omega,t}.\label{E25}
\end{align}

In order to bring Eqs.~\eqref{E24} and \eqref{E25} into a convenient form, several steps have to be performed, which are outlined in appendix \ref{AS1}. First, the Green's functions and the occupation function have to be expressed in the Dirac interaction picture. This guarantees that the l.h.s. of \eqref{E24} becomes proportional to the temporal derivative of $F$. Second, one has to apply the Wigner approximation to the second term of the collision integral \eqref{E25}. The latter approximates convolution with respect to frequency and time in Wigner representation by the product 
\begin{align}
I^{\text{coll}}_{p,\omega,t}\approx\Sigma^{K}_{p,\omega,t}-(\Sigma^{R}_{p,\omega,t}\Sigma_{z}F_{p,\omega,t}-F_{p,\omega,t}\Sigma_{z} \Sigma^{A}_{p,\omega,t}).\label{WigA}
\end{align}
As detailed in the appendix and Refs.~\cite{buchhold2015nonequilibrium,buchhold2015kinetic,buchhold2016prethermalization}, this is justified as long as the condition 
\begin{align}
\frac{|\partial_tF_{\alpha\beta,q,\omega,t}||\partial_{\omega}\Sigma^R_{\alpha\beta,q,\omega,t}|}{|F_{\alpha\beta,q,\omega,t}||\Sigma^R_{\alpha\beta,q,\omega,t}|} \ll 1\label{E27}
\end{align}
is fulfilled. Physically, Eq.~\eqref{E27} requires that the characteristic timescale of the global system's time evolution is much larger than the characteristic time scale of a single scattering event. This is justified for splitting protocols, at sufficiently low energies, for which (see Ref.~\cite{buchhold2015kinetic})
\begin{align}
\epsilon v\ll \nu^2,
\end{align}
where $\epsilon$ is the average energy per momentum mode in the system.\\

Multiplying both sides of the kinetic equation \eqref{E24} with the system's spectral functions
\begin{align}
\mathcal{A}_{p,\omega,t}=\frac{i}{2\pi}(G^{R}_{p,\omega,t}-G^{A}_{p,\omega,t}),\label{E28}
\end{align}
projects the kinetic equation onto the region where $\mathcal{A}_{p,\omega,t}$ has non-zero support, i.e. onto the physical quasi-particle states. For a given momentum and energy $p, \epsilon_p$ the width of the spectral function can be estimated to be $\sim \sigma^R_p\ll\epsilon_p$ in the relevant regime of low momenta and small energy density \cite{buchhold2015kinetic,punk2006collective}. The corresponding criterion in the present setting is identical to the criterion for Luttinger theory to be applicable, i.e. to have a sufficiently low excitation energy density $n_q<\Lambda/q$ (cf. \cite{buchhold2015nonequilibrium}).
The kinetic equation is thus projected on-shell and 
one can make use of Eqs.~\eqref{E22} and \eqref{E23} to simplify the collision integral $I^{\text{coll}}$.
On the other hand (for $\alpha,\beta=1,2$) \eqref{Partnum1}, 
\begin{align}
\int_{\omega}\left(\mathcal{A}\Sigma_{z}F\Sigma_{z}\right)_{\alpha\beta,p,\omega,t}\approx&
\begin{pmatrix}
  2n_{\alpha\beta,p,t}+\delta_{\alpha,\beta} & - 2m_{\alpha\beta,p,t} \label{E29}\\ 
   -2m_{\alpha\beta,p,t} &  2n_{\alpha\beta,p,t}+\delta_{\alpha,\beta}
 \end{pmatrix},
\end{align}
which finally leads to
\begin{align}
\partial_tn_{11,p,t}&=\sigma^K_{11,p,t}-\sigma^R_{p,t}(2n_{11,p,t}+1),\ \label{E30}
\end{align}
\begin{align}
\partial_tn_{12,p,t}&=\sigma^K_{12,p,t}-2\sigma^R_{p,t}n_{12,p,t},\ \label{E31}\\
\partial_tm_{11,p,t}&=\Phi^K_{11,p,t}-2\sigma^R_{p,t}m_{11,p,t}.\label{E32}
\end{align}
The anomalous densities $m_{12,p,t}$ never deviate from zero and their time evolution is not considered here.

The self-energies are evaluated in self-consistent Born approximation, see appendix \ref{AS3}. 
An additional rescaling of time $\tilde{t}=vt$ and the retarded self-energy $\tilde{\sigma}^{R}_{p,t}=v^{-1}\sigma^{R}_{p,\tilde{t}}$ removes the explicit dependence of $\tilde{\sigma}^R_{p,\tilde{t}}$ and $n_{p, \tilde{t}}, m_{p,\tilde{t}}$ on the non-linearity $v$ completely. This yields the equations of motion for the phonon densities
\begin{widetext}
\begin{align}
\partial_{\tilde{t}}n_{11,p}&=\int_{0<q<p}\hspace{-0.5cm}\gamma^{-}_{p,q}[n_{11,p-q}n_{11,q}-n_{11,p}(n_{11,p-q}+n_{11,q}+1)]\label{E33}+2\int_{0<q}\hspace{-0.2cm}\gamma^{+}_{p,q}[n_{11,p+q}(1+n_{11,q}+n_{11,p})-n_{11,q}n_{11,p}],\\
\partial_{\tilde{t}}n_{12,p}&=\int_{0<q<p}\hspace{-0.5cm}\gamma^{-}_{p,q}[n_{12,p-q}n_{12,q}-n_{12,p}(n_{11,p-q}+n_{11,q}+1)]\label{E34}+2\int_{0<q}\hspace{-0.2cm}\gamma^{+}_{p,q}[(n_{11,p+q}-n_{11,q})n_{12,p}-n_{12,p+q}n_{12,q}],\\
\partial_{\tilde{t}}m_{11,p}&=\int_{0<q<p}\hspace{-0.5cm}\gamma^{-}_{p,q}[m_{11,p-q}m_{11,q}-m_{11,p}(n_{11,p-q}+n_{11,q}+1)] \label{E35}+2\int_{0<q}\hspace{-0.2cm}\gamma^{+}_{p,q}[(n_{11,p+q}-n_{11,q})m_{11,p}-m_{11,p+q}m_{11,q}],\ \ \ \ \ \ \ \ \ \ \ 
\end{align}
\end{widetext}
with the effective scattering vertex 
\begin{align}
\gamma^{\pm}_{p,q}=\frac{qp(p\pm q)}{\tilde{\sigma}^{R}_{p}+\tilde{\sigma}^{R}_q+\tilde{\sigma}^{R}_{p\pm q}}.\label{VX}
\end{align}

\begin{figure}
	\includegraphics[width=\linewidth]{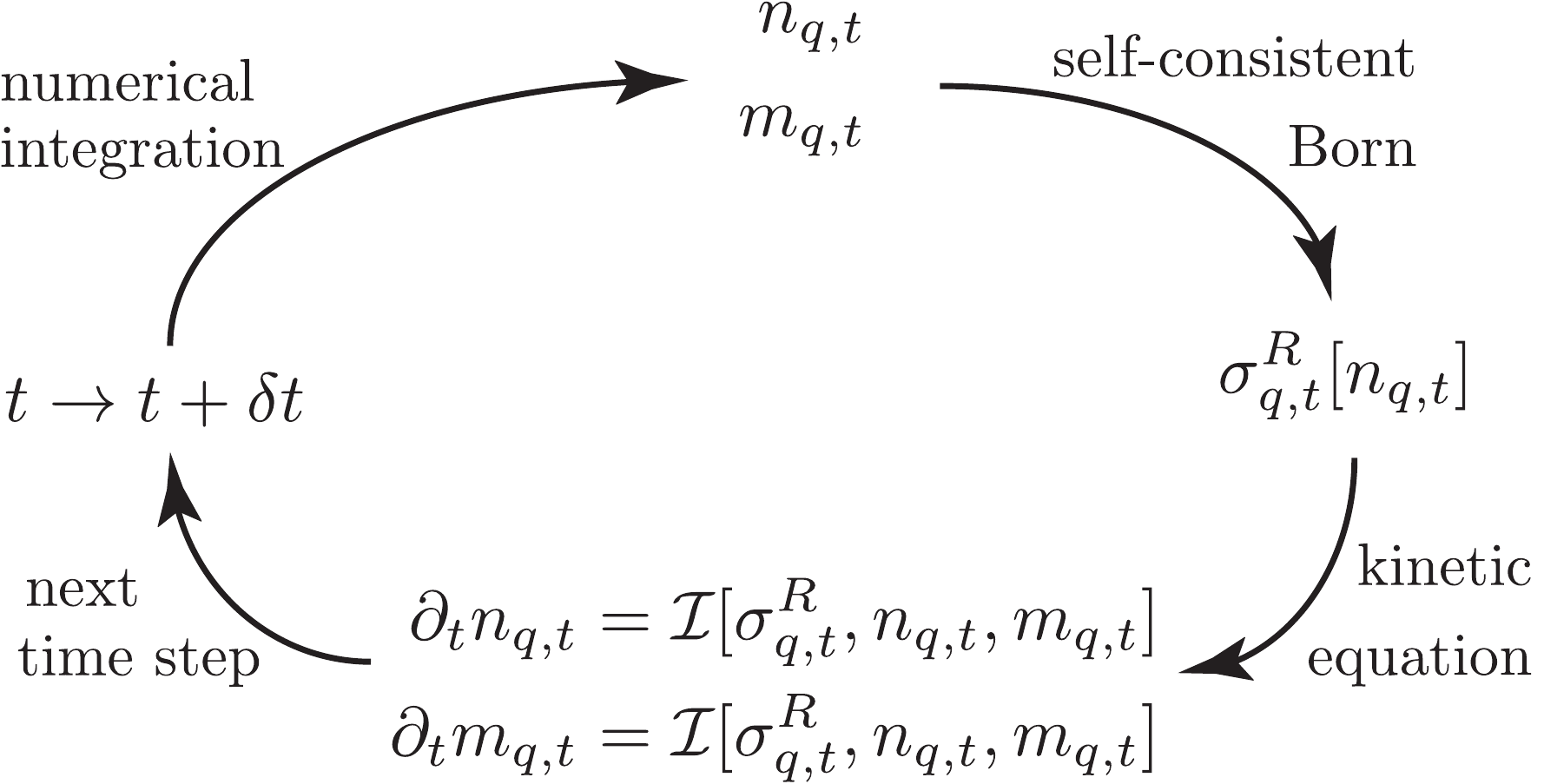}
	\caption{Schematic illustration of the numerical routine for solving the kinetic equations. For a given occupation $n_{q,t}, m_{q,t}$, the self-energies are computed within self-consistent Born approximation. Inserting the self-energies into the collision integral determines the evolution of $n_{q,t}, m_{q,t}$ via the kinetic equation. Numerical integration yields the occupations at the next time step $t\rightarrow t+\delta t$.}
	\label{F1}
\end{figure}

At each time step, the retarded self-energy has to be determined self-consistently. It is determined by the equation (see appendix \ref{AS3})
\begin{align}
\tilde{\sigma}^{R}_p=2\int_{q}\Big[&\left(\frac{\partial_{\tilde{t}}\ n_{11,q}}{\tilde{\sigma}^{R}_q}+(2n_{11,q}+1)\right)\nonumber\\ &\left(\frac{qp(p-q)}{\tilde{\sigma}^{R}_q+\tilde{\sigma}^{R}_{p-q}}+\frac{qp(p+q)}{\tilde{\sigma}^{R}_q+\tilde{\sigma}^{R}_{p+q}}\right)\Big].\ \label{E36}
\end{align}
The system of equations \eqref{E33}-\eqref{E36} is solved iteratively according to the scheme shown in Fig.~\ref{F1}. The initial state at $\tilde{t}=0$ is determined by the individual splitting procedure. At each time step $\tilde{t}$, one has to compute the self-energy $\tilde{\sigma}^R_{p,\tilde{t}}$ which then enables the next infinitesimal time step $\tilde{t}+\delta$ via the equations of motion. The procedure is repeated until the steady state has been reached.

To obtain the closed set of Eqs.~\eqref{E33}-\eqref{E36}, the Dyson-Schwinger equations for the interacting Luttinger liquid have been truncated at the quadratic order, i.e. we neglect any vertex correction to Eq.~\eqref{VX} from higher order loops. In principle it is possible to include these corrections within the present formalism as was discussed in Ref.~\cite{buchhold2015kinetic}. This work shows, however, that higher order vertex corrections are generally very small, which is in agreement with numerical test simulations of the vertex correction for the present setup, and can thus be safely neglected. At this order, the Dyson-Schwinger equations are identical to the self-consistent Born approximation.

\section{Splitting to a nearly thermal state}\label{A3}
The splitting experiments in the Vienna group have been performed via two different splitting protocols. In this section, we will investigate a set of splitting experiments, in which the bosonic wires were split in such a way that the resulting state of the system remained close to a thermal state. We demonstrate that, as a consequence, the prethermalized state and the asymptotic thermal state are hard to distinguish and the asymptotic thermalization dynamics is difficult to track experimentally. We will discuss the different stages of thermalization for this splitting procedure and numerically investigate the behavior of observables in the corresponding stages. This serves as a preparation for the discussion of the second set of splitting procedures in the subsequent section.
\subsection{Initial state}
The initial state in the two wires is prepared by coherently splitting a phase-fluctuating one-dimensional quasi-condensate in the longitudinal direction. The splitting is performed on a fast time scale such that the longitudinal phase profiles in the newly generated quantum wires are nearly identical $\theta_1(x)\approx \theta_2(x)$ \cite{kitagawa2011dynamics,geiger2014local,gring2012relaxation}, while the uncertainty in the local density is maximal.

We label the phase and density $\theta_\alpha, \rho_\alpha$ of the two wires with $\alpha=1,2$ and introduce the relative $\theta_r=\theta_1-\theta_2$ and absolute phase $\theta_a=\frac{\theta_1+\theta_2}{2}$ and analogously relative and absolute density $\rho_r, \rho_a$. The absolute phase and density remain unchanged by a sudden splitting and correspond to the initial quasi-condensate. The relative phase and density instead are generated only by the splitting. Their initial variance can be expressed via the initial average density in the wires $\rho_0$ and per unit volume reads as \citep{kitagawa2011dynamics,gring2012relaxation,geiger2014local}
\begin{align}
\langle \rho_{r}(x)\rho_{r}(x) \rangle &\approx \frac{\rho_{0}}{2},\label{E13}\\
\langle \theta_{r}(x)\theta_{r}(x) \rangle &\approx \frac{1}{2\rho_{0}}.\label{E14}
\end{align}

The quasiparticle densities in the relative mode can be determined from Eqs.~\eqref{E13}-\eqref{E14} in combination with Eqs.~\eqref{E6}-\eqref{E7},
\begin{align}
 n_{r,p}&=\langle a^{\dagger}_{r}(p)a_{r}(p)\rangle\ =\frac{T_{r}}{\nu |p|}+\chi_{r}|p|-\frac{1}{2},\label{E15}\\
 m_{r,p}&=\langle a_{r}(p) a_{r}(-p)\rangle=\frac{T_{r}}{\nu |p|}-\chi_{r}|p|.\label{E16}
\end{align}
At low momenta $|p|<\sqrt{\tfrac{T_r}{\nu\chi_r}}$ the normal density adopts thermal behavior $n_{r,p}\sim |p|^{-1}$ with an effective relative temperature $T_{r}= \frac{\rho_{0}\pi}{4}\frac{\nu}{K}$, which depends on the state before the splitting via $\rho_0$, and on the Hamiltonian via the Luttinger parameters. For intermediate momenta $\sqrt{\tfrac{T_r}{\nu\chi_r}}<|p|<\Lambda$, on the other hand, it increases linearly in momentum with a splitting dependent slope $\chi_{r}=\frac{1}{4\pi\rho_{0}}K$. On the largest momenta $|p|>\Lambda$ the quasiparticle density is suppressed exponentially. Analogous behavior is found for the anomalous density $m_{r,p}$, which must, however, be zero for a thermal ensemble. This should be contrasted with the thermal occupation of the absolute quasiparticle mode  \citep{kitagawa2011dynamics,gring2012relaxation,geiger2014local}
\begin{align}
 n_{a,p}&=\langle a^{\dagger}_{r}(p)a_{r}(p)\rangle =\frac{\cosh(\nu|p|/T_a)-1}{2} \approx\frac{T_{a}}{\nu |p|},\label{E15a}\\
 m_{a,p}&=\langle a_{a}(p) a_{a}(-p)\rangle=0,\label{E16a}
\end{align}
which is described by the initial temperature $T_a$ of the quasi-condensate before the splitting.

The low momentum effective temperature of the relative mode $T_r\sim \rho_0/K$ can be initialized at rather small values, depending on the ratio of the initial density and the Luttinger parameter. On the other hand, decreasing $T_r$ increases the slope $\chi_r\sim K/\rho_0$ and shifts the splitting energy from lower to larger momenta. The larger $\chi_r$, the more $n_{r,p}$ deviates from a thermal distribution and the larger will be the impact of energy redistributing collisions at large times. In previous experiments, the linear increase of the occupation has not been detected in any of the observables, indicating that the crossover momentum $\sqrt{\frac{T_r}{\chi_r\nu}}$ is large, i.e. of the order of the ultraviolet cutoff and thus $\chi_r$ is small. While this still implies an energy transfer from the high momentum to the low momentum regime on times of the order of the collision time, our results show that the initial, thermal form of the low momentum distribution makes such transfer hard to detect.
This is discussed in detail in Secs.~\ref{A5} and \ref{A7}.  An illustration of the relative and absolute densities is shown in Figs.~\ref{F4} and~\ref{F7} for two different sets of splitting parameters.

\begin{table}
  \centering
   \begin{tabular}{| l || l |}
   \hline
   Experimental parameters &  Luttinger liquid parameters\\
    \hline
    $m_{\text{Rb}} =87u$&$K  = 25.4$ \\
    $\rho_{0} = 30 \cdot 10^{6} \text{m}^{-1}$ & $\nu = 1.3 \cdot 10^{-3} \tfrac{\text{m}}{\text{s}}$\\
    $T_{a} = 30 \,\text{nK}$ & $v =5.4 \cdot10^{-10}\tfrac{\text{m}^2}{\text{s}}$\\
    $u(0) \approx 8.8 \cdot 10^{-39} \text{Jm}$ & $T_{r} = 21.5 \,\text{nK}$\\
  \phantom{$u(0) \approx 8.8 \cdot 10^{-39} \text{Jm}$}& $\Lambda = 2.5\cdot 10^{6}\text{m}^{-1}$\\
  \hline
  \end{tabular}
    \caption{Left column: Microscopic parameters, which are used throughout this work in order to perform the numerical simulations. They represent a set of experimentally realizable conditions and can be found e.g. in Refs.~\citep{gring2012relaxation,kitagawa2011dynamics,monien1998trapped,langen2013local,olshanii1998atomic}. $m_{\text{Rb}}$ is the mass of a $\text{Rb}^{87}$ atom, which are used in the experiment and $u=1.66\cdot 10^{-27}$kg is the atomic mass unit. For typical experiments, the relative and absolute temperature vary between $10\text{nK}\le T_a\le 100\text{nK}$ and $5\text{nK}\le T_r\le25\text{nK}$.
    Right column: Luttinger liquid parameters inferred from the microscopic parameters.}
 \label{Tabel1}
\end{table}

As pointed out in the previous section, the kinetics of the quasiparticles is most conveniently expressed in the basis of individual wires. Inserting the definition of the relative and absolute mode $a_{r,a}=\tfrac{1}{\sqrt{2}}(a_1\mp a_2)$ into the densities one finds immediately
\begin{align}
n_{r,p}=&n_{11,p}-n_{12,p},\label{E17}\\
m_{r,p}=&m_{11,p},\label{E18}\\
n_{a,p}=&n_{11,p}+n_{12,p},\label{E19}
\end{align}
where we used the symmetry $n_{22,p}=n_{11,p}$ and $n_{12,p}=n_{21,p}$ between the two subsystems and $m_{a,p}=0\Leftrightarrow m_{11}=-m_{12}$. The initial values can be read off directly from Eqs.~\eqref{E15}-\eqref{E16a}.
\begin{figure*}
\centering
  \includegraphics[width=\linewidth]{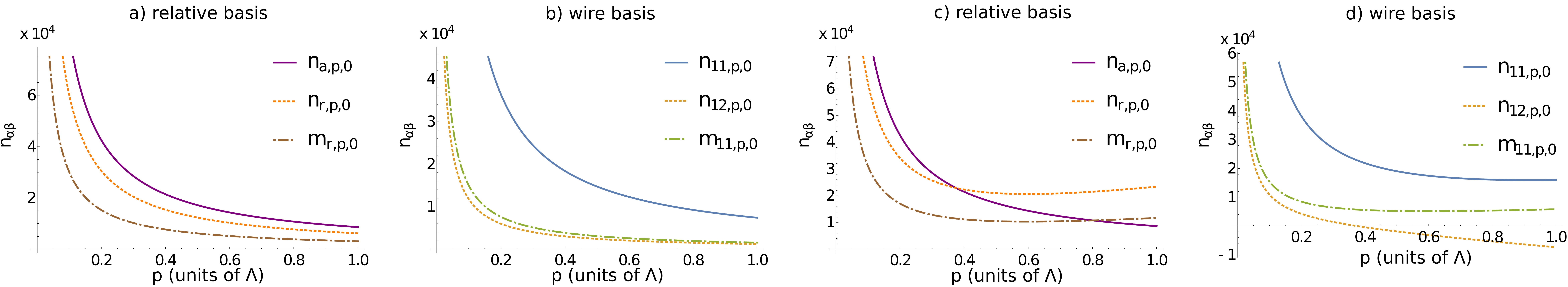}
    \caption{For parameters as in Table~\ref{Tabel1} but $\chi_r$ set to zero $\chi_r=0$ by hand: a) initial distribution in the absolute $n_a$ and relative $n_r, m_r$ modes, b) initial distribution in the single wire basis. We focus in this analysis on the low-energy dynamics of the system below a certain cutoff $p<\Lambda$. Due to equations \eqref{E15} and \eqref{E16}, the initial distributions are almost thermal for all momenta.\\
For parameters as in Table~\ref{Tabel1} but $\chi_r=\frac{1}{4 \pi \rho_{0}}K$:
c)+d) The presence of $\sqrt{T_r/(\nu \chi_r)}\approx 0.5\Lambda<\infty$ distinguishes two momentum regimes: Small momenta $|p|<\sqrt{T_r/(\nu \chi_r)}$ with a thermal relative density $n_r=T_r/(\nu |p|)<n_a$ and large momenta $|p|>\sqrt{T_r/(\nu \chi_r)}$ with a linear increasing density $n_r=\chi_r |p|>n_a$.}\label{F2}
\end{figure*}

\subsection{Prethermalization and Thermalization: Time regimes in the double-wire dynamics}\label{A5}
The splitting procedure initializes the two-wire system in a non-thermal, out-of-equilibrium state, which will start to relax immediately after the splitting. This relaxation process undergoes the typical, stepwise process towards equilibration, which we will discuss in the following. We start with a qualitative discussion of the different relaxation processes based on the Hamiltonian \eqref{Ham8} and the initial state properties, expressed via $n_{11}, n_{12}, m_{11}$. This scenario is then confirmed quantitatively by simulations of the kinetic equations \eqref{E33}-\eqref{E35} and discussed on the basis of experimentally relevant observables. 
The relaxation of a generic, weakly interacting quantum system is expected to happen in a sequence of steps connecting the initial non-equilibrium state with the asymptotic state of the system \cite{kollar2011generalized, buchhold2016prethermalization, moeckel2008interaction}. We will review this typical scenario briefly, adapted to the present system.

Straight after the initialization of the out-of-equilibrium state, the time evolution of the system is dominated by the fastest process captured by the Hamiltonian \eqref{Ham8} with $v=0$, which is the formation of well-defined, weakly interacting quasiparticles. Typical for this first process is the dephasing of  anomalous correlations 
\begin{align}
\langle a_{\alpha,p}a_{\beta,-p}\rangle_t\approx e^{i2\nu|p|t}\langle a_{\alpha,p}a_{\beta,-p}\rangle_{t=0},
\end{align}
caused by the quadratic part of the Hamiltonian. The spatio-temporal regime, which is influenced by the dephasing is spanned by a pair of counter propagating quasiparticles, leading to a light cone whose shape is set by the phonon velocity. Outside this light cone, correlations are still determined by the initial state at $t=0$. We term this regime "initial state regime". Inside the light cone quasiparticles have formed and observables are dominated by the time-dependent quasiparticle densities $n_{11,p,t}, n_{12,p,t}$, whereas the off-diagonal part $m_{11,p,t}$, due to the dephasing, is effectively zero. 

As long as both wires are described by the same microscopic parameters, the densities $n_{11,p,t}, n_{12,p,t}$ are constants of motion of the quadratic Luttinger Hamiltonian. Their time evolution is solely set by energy redistributing collisions as described in the nonlinear part of $H_{\text{ILL}}$. Since the quasiparticle interactions are weak and display subleading scaling behavior in momentum space, the typical time scale for long wavelength collisions is much larger than for the dephasing, i.e. for a momentum mode $q$,  $t_{\text{coll}}=(v|q|^\frac{3}{2}(T_r/\nu)^{\frac{1}{2}})^{-1}\gg(\nu|q|)^{-1}=t_{\text{deph}}$ \cite{buchhold2015nonequilibrium,buchhold2016prethermalization}. Consequently, there exists an extended spatio-temporal regime for which dephasing has lead to the formation of quasiparticles, which have, however, not experienced a substantial number of collisions. In this regime, the densities $n_{11,p,t}, n_{12,p,t}$ are pinned to their initial values and are not yet evolving in time. The corresponding state is commonly termed prethermal.

For times $t>t_{\text{coll}}$, typically only few collisions ($\sim 3$) are necessary to redistribute energy and establish a local thermal equilibrium. By local, we refer to a local region in momentum space $p\in [q-\delta, q+\delta]$, which is very well described by thermal quasiparticle densities $n_{12,p,t}=0, n_{11,p,t}=n_{B}(\nu |p|/T_{\text{loc}})$ and a joint, local effective temperature $T_{\text{loc}}$. This effective thermalization time scale is of the order of few collision time scales $t_{\text{therm}}=\mathcal{O}(t_{\text{coll}})$ and therefore, together with $T_{\text{loc}}$, momentum dependent. 

On asymptotic time scales, the locally thermalized regions have to adjust their temperature and relax towards the global thermal equilibrium. At this stage of the relaxation, conserved quantities like energy and momentum have to be transferred over large distances in momentum space. This transport involves many subsequent collisions and belongs to the regime of hydrodynamics, which is governed by the global conservation laws of the system. This regime makes a clear distinction between the inter-wire coherences $n_{12,p,t}$, which are not part of any conservation law and thus decay exponentially towards their thermal value, and the intra-wire density $n_{11,p,t}$, whose evolution is constrained by total energy and momentum conservation. As a consequence, $n_{11,p,t}$ will relax algebraically slowly towards its thermal value in this asymptotic regime. 

In the following, the time evolution of the densities $n_{11,p,t}, n_{12,p,t}, m_{11,p,t}$ is determined on the basis of the kinetic equations~\eqref{E33}-\eqref{E35} and the emergence of the different relaxation regimes is discussed on the basis of several observables.

\begin{figure}
	\includegraphics[width=\linewidth]{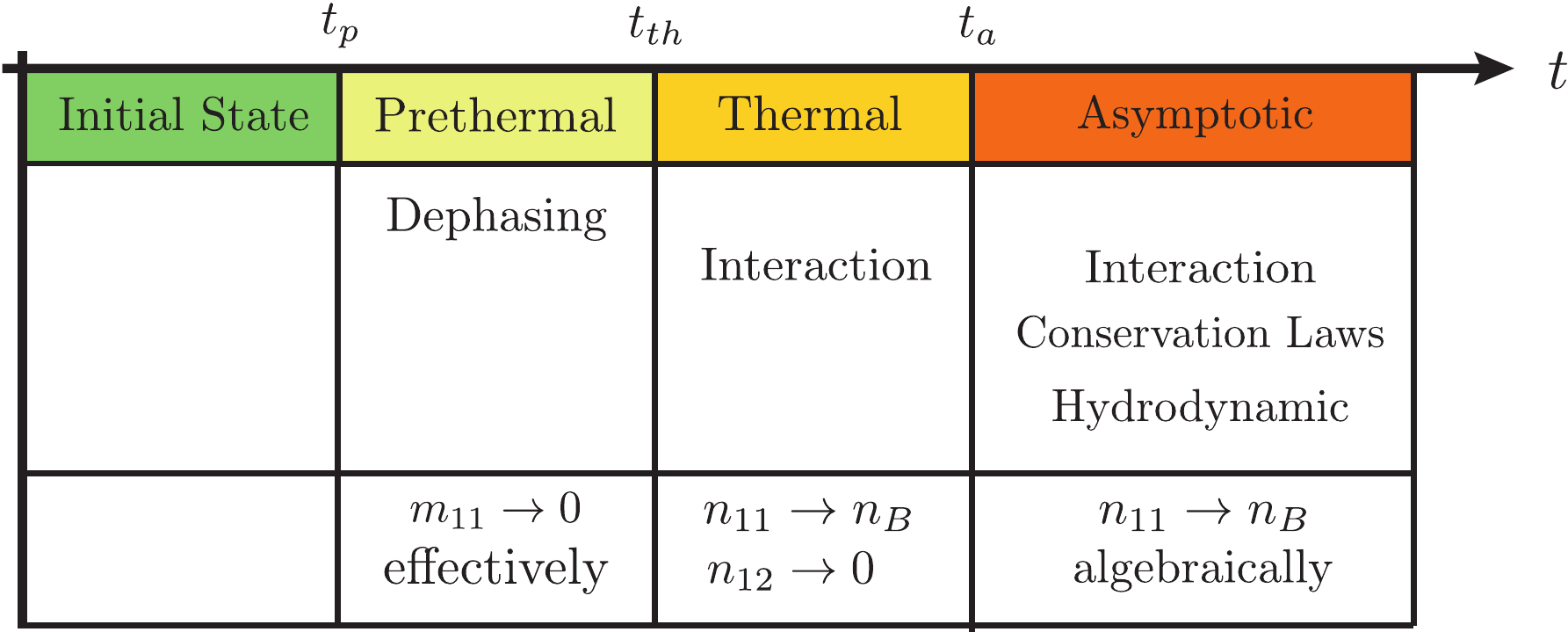}
	\caption{Illustration of the various stages of thermalization for an arbitrary but fixed patch in momentum space centered around momentum $q$. After an initial state dominated time interval $0\le t<t\sub{deph}=(\nu|q|)^{-1}$, the dephasing of the anomalous modes $m_{11,q}$ becomes significant, introducing a prethermal regime. Collisions between the phonon modes modify the normal densities $n_{11,q}, n_{12,q}$ on time scales $t=t\sub{coll}=(v|q|^\frac{3}{2}(T_r/\nu)^{\frac{1}{2}})^{-1}$ and bring the system into a thermalizing regime for times $t>t\sub{therm}=\mathcal{O}(t\sub{coll})$. The thermalizing regime is separated into an initial stage, governed by a fast evolution of the normal densities due to few, local collisions and an asymptotic stage, governed by energy and momentum transport over many distinct momentum patches. This stage involves many collisions, captured by slow, hydrodynamic modes.}
	\label{F5}
\end{figure}
\begin{figure*}
\centering
  \includegraphics[width=\linewidth]{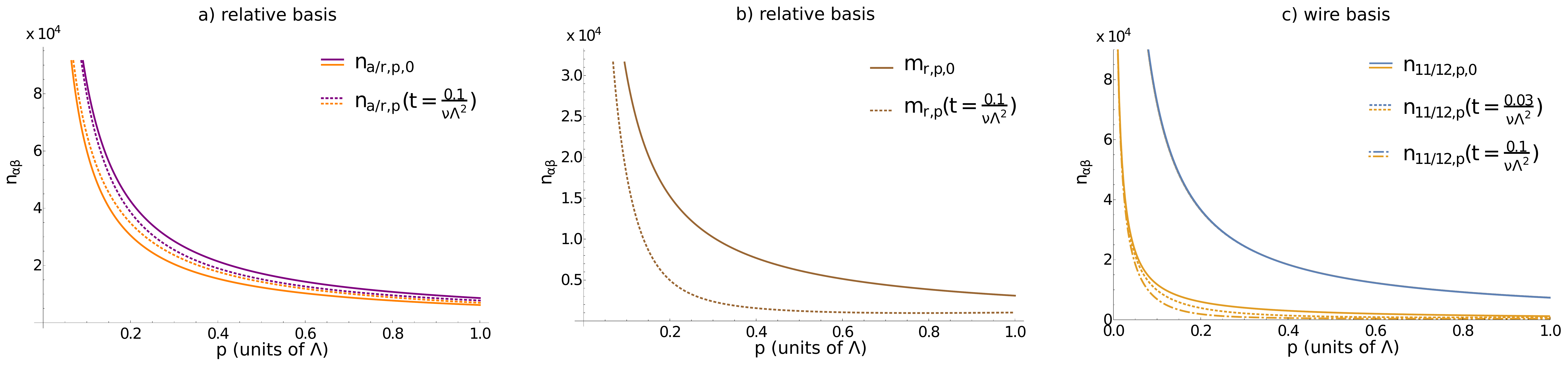}
\caption{Parameters as in Table~\ref{Tabel1} but $\chi_r=0$: a) The relative and absolute phonon densities, $n_{i,p}$ with $i\in \{r,a\}$, with an initial thermal distribution (bold lines) as shown in Fig.~\ref{F2} approach each other as time evolves. Both occupations converge against a final thermal state characterized by a single temperature $T\sub{f}$. b) The anomalous density $m_{r,p,t}$ decays in time and approaches zero. c) In this case, $\chi_r\approx0$ is negligible and the evolution is governed by the slow decay of $n_{12,p,t}$ in time, while $n_{11,p,t}$ remains constant in time (all blue (dark) lines coincide).}
\label{F4}
\end{figure*}
\subsubsection{Time-evolution of the phonon densities}
The kinetic equations~\eqref{E33}-\eqref{E35} determine the relaxation of the densities $n_{11,p,t}, n_{12,p,t}$ and $m_{11,p,t}$ towards a thermal ensemble. The Hamiltonian \eqref{Ham3} contains no interactions between the two wires and between states of opposite momentum, such that the off-diagonal terms are expected to decay to zero $n_{12,p,t\rightarrow\infty}=m_{11,p,t\rightarrow\infty}=0$ for a thermal state. The different stages of relaxation are as follows.

For times $t<t\sub{therm}$, the normal densities $n_{11,p}, n_{12,p}$ are nearly constant and the only relaxation is caused by the dephasing of $m_{11,p}$. This time regime is adequately described and understood in terms of the quadratic Hamiltonian alone. On the other hand, for times $t>t\sub{therm}$, the dephasing of the anomalous modes has already eliminated all the influence of $m_{11,p}$ on physical observables, although its absolute value remains non-zero. As it turns out, the dynamics for $t>t\sub{therm}$ can be understood completely on the basis of the relaxation of the normal occupations, and we will thus restrict the following discussion on the time evolution of $n_{11,p,t}, n_{12,p,t}$. The simulated relaxation of $m_{11,p,t}$ is, however, shown for completeness. 

The diagonal density $n_{11,p,t\rightarrow\infty}=n\sub{B}(\epsilon_p/T\sub{f})$ evolves towards a Bose-Einstein distribution $n\sub{B}$ with asymptotic temperature $T\sub{f}$. In the basis of relative and absolute modes, both normal occupations approach each other during the evolution and converge towards an identical, thermal distribution $n_{r,p,t\rightarrow\infty}=n_{a,p,t\rightarrow\infty}=n\sub{B}(\epsilon_p/T\sub{f})$, as can be inferred from Eqs.~\eqref{E17}-\eqref{E19}. The anomalous occupation $m_{r,p,t\rightarrow\infty}=0$ instead decays to zero. 

The asymptotic value of the temperature of the diagonal, normal phonon distribution $n_{11,p,t\rightarrow\infty}=n\sub{B}(\nu|p|/T\sub{f})$ can be determined by the initial state due to energy conservation. The quadratic part of the Hamiltonian, i.e. the kinetic energy density, commutes with the resonant collision terms in Eq.~\eqref{Ham8}. The kinetic energy density is thus an integral of motion and its final and initial value must coincide $\epsilon_{t=0}=\epsilon_{t=\infty}$, i.e. 
\begin{equation}\label{asympT}
\int_p \nu |p|n_{11,p,t=0}\overset{!}{=}\int_p\nu |p| n_{11,p,t=\infty}=\frac{\pi^2 T\sub{f}^2}{3\nu}.
\end{equation}
This determines the asymptotic temperatures $T\sub{f}=\sqrt{3\nu\epsilon_{t=0}}/\pi$.

For initial states of the form given in Eq.~\eqref{E15}, one can distinguish between two experimentally relevant limiting cases, namely $K\ll \rho_0 \Leftrightarrow \chi_r\approx 0$, where $\chi_r \Lambda$ is negligibly small, and $\chi_r\Lambda\approx T_r/(\nu\Lambda)$, for which the linear tail $\sim\chi_rp$ cannot be neglected. 

In the former case, the relative and absolute mode are both described by an approximate high temperature distribution $n_{\alpha,p}\approx T_{\alpha}/(\nu |p|)$, where $\alpha=a,r$. The time evolution then lets both distributions approach each other slowly, merging at a joint temperature $T\sub{f}=(T_r+T_a)/2$. In the basis of $n_{11,p}, n_{12,p}$, such initial states lead to an almost time independent diagonal occupation $n_{11,p,t}=n_{11,p,0}$ and only $n_{12,p,t}$ is decaying to zero. The results of a corresponding kinetic equation simulation are shown in Fig.~\ref{F4}.

\begin{figure*}
\centering
   \includegraphics[width=\linewidth]{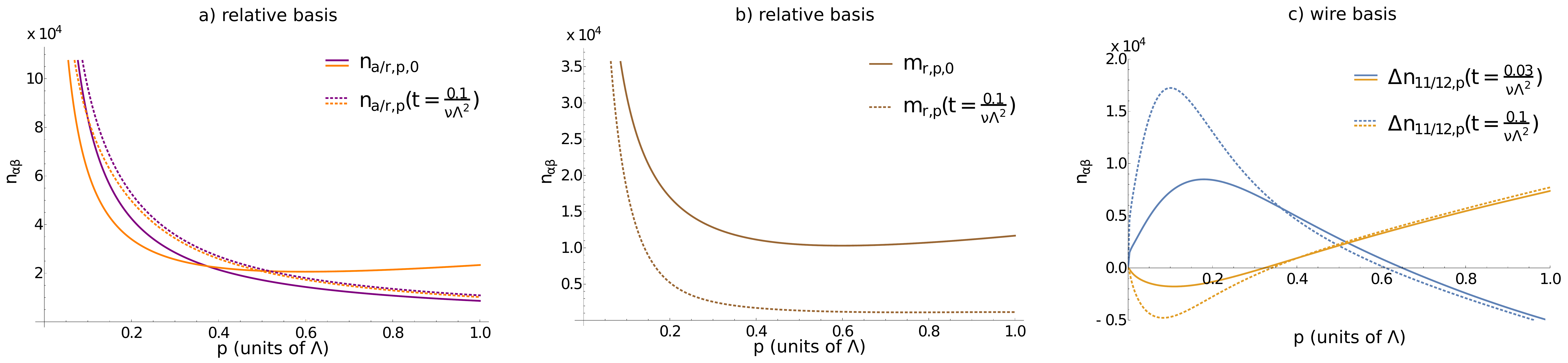}
\caption{Parameters as in Table~\ref{Tabel1} with $\chi_r=\frac{1}{4 \pi \rho_{0}}K$: a) The initial diagonal relative phonon density $n_{r,p}$ (bright, bold) has a linear tail (see Fig. \ref{F2}) and energy stored at larger momenta. During the time evolution, $n_{r,p,t}$ and $n_{a,p,t}$ converge towards an identical final distribution, which corresponds to a larger temperature. Energy is distributed from larger to small momenta. b) The off-diagonal occupation decays to zero exponentially in time. c) Time evolution of the difference $\Delta n_{\alpha, p}(t)=n_{\alpha,p}(t)-n_{\alpha,p}(0)$. The crossover momentum $p^{(c)}_t$ separating fast, thermal-like evolution at $p>p^{(c)}_t$ from a slow non-equilibrium relaxation at $p<p^{(c)}_t$ leads to a bump in $\Delta n_{\alpha, p}(t)$ as discussed in the main text.}
\label{F7}
\end{figure*}

In the case of $\chi_r\neq0$ or generally for a relative distribution $n_{r,p,t}$ deviating from a thermal one, not only $n_{12,p,t}$ decays in time, but also $n_{11,p,t}$ will evolve in a non-trivial way. Since the Hamiltonian conserves the energy within a single wire, $n_{11,p,t}$ is not simply decaying but energy is redistributed within this mode. For an initial distribution of the form \eqref{E15} with $\chi_r\neq0$, energy is initially stored at large momenta $ \sqrt{T_r/(\nu \chi_r)}<q<\Lambda$ and will be transferred to lower momenta via phonon-phonon collisions in order to reach a thermal state. In Fig.~\ref{F7}, the phonon distributions for such a scenario are shown. One can clearly observe the time evolution of the diagonal density $n_{11,p,t}$ and its effect on $n_{r,p,t}, n_{a,p,t}$, which are both evolving towards an increased temperature state with $T\sub{f}>(T_r+T_a)/2$.

Due to the structure of the interaction vertex $V(p,q,p\pm q)=v\sqrt{|pq(p\pm q)|}$ the collision rate decreases with the momentum of the contributing phonon modes. The modification of $n_{\alpha,p,t}$ due to collisions sets in at large momenta and approaches lower and lower momenta as time proceeds.  In Fig.~\ref{F7} c), the deviation $\Delta n_{\alpha,p,t}=n_{\alpha,p,t}-n_{\alpha,p,0}$ of the densities $\alpha=11, 12$ from their initial values is shown for different time steps. It has been observed previously \cite{buchhold2015nonequilibrium,buchhold2016prethermalization} that this deviation displays two characteristic momentum regimes, a low momentum regime $p<p^{(c)}_t$ with linear deviation $\Delta n_{\alpha,p,t}\sim |p|$ and a large momentum regime $p>p^{(c)}_t$ with a Rayleigh-Jeans type deviation $\Delta n_{\alpha,p,t}\sim 1/|p|$. In this large momentum regime, the phonon modes have reached a local, nearly thermal equilibrium and $\Delta n_{\alpha,p,t}$ evolves very slowly in time. In the vicinity of the crossover momentum $p^{(c)}_t$, however, the redistribution of energy is fast and $\Delta n_{\alpha,p,t}$ undergoes a strong evolution. The time evolution of $p^{(c)}_t\sim t^{-\alpha}$ is algebraic in time with $1/2\le\alpha<1$ depending on the initial state \cite{buchhold2016prethermalization}. For a nearly thermal initial state $\alpha\approx2/3$, which is consistent with the typical collision time $t\sub{coll}=(v|q|^{\frac{3}{2}}(T_r/\nu)^{\frac{1}{2}})^{-1}$. 
Tracking the evolution of the crossover momentum $p^{(c)}_t$ in the experiment can thus be used to detect the thermal relaxation exponent $\alpha=2/3$ predicted for the one-dimensional Bose gas \cite{buchhold2015nonequilibrium,buchhold2016prethermalization,andreev1980hydrodynamics,punk2006collective}.

\subsubsection{Dynamics of the relative phase correlation function}\label{A7}
The phase correlation function $C(x,t)\sim \exp(-\Delta\theta_{r,x,t}/2)$ defined in Eqs.~\eqref{E47}, \eqref{E48} is an experimentally accessible measure to monitor the time evolution of the two quantum wires \cite{kitagawa2011dynamics,geiger2014local,gring2012relaxation,langen2015experimental}. It has been discussed extensively on the level of the quadratic Luttinger model without phonon collisions, which describes very well the dynamics of the initial and the prethermal state. Below, we will discuss in which way the phase correlation function is modified by phonon-phonon collisions. We show that $C$ is dominated by the thermal part of the phonon distribution and that it is quite insensitive towards a modification of $n_{r,p,t}$ of the form $\Delta_p=\chi_r|p|$. As a consequence, the thermalization dynamics of $C$ for an initial state of the form \eqref{E15} does not display any pronounced additional features compared to a collisionless evolution. The latter signals the evolution of the system towards a GGE state caused by the dephasing of $m_{p,t}$. The only resolvable effect of the collisions is an effective temperature in the relative mode $T\sub{eff}>T_r$, slightly larger than the initial state $T_r$. 

\begin{figure*}
	\includegraphics[width=\linewidth]{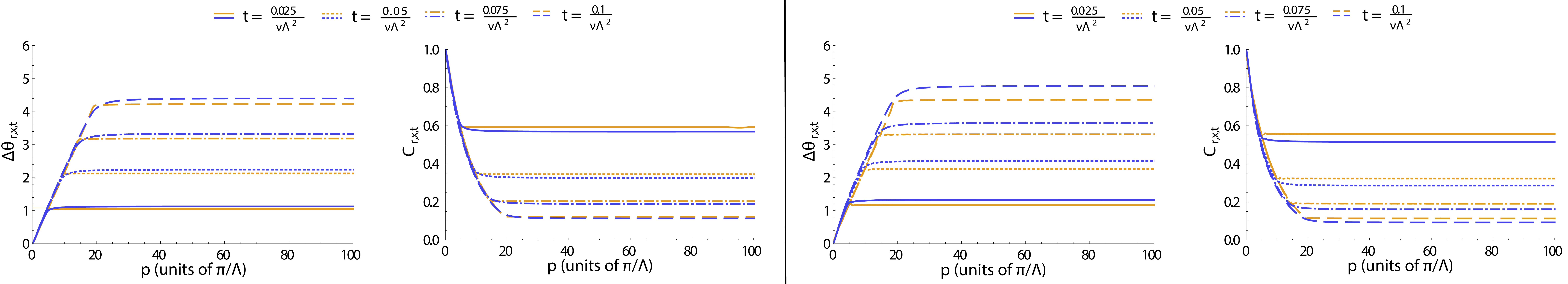}
	\caption{Relative phase fluctuation $\Delta\theta_r$ and phase correlation function $C$ for a nearly thermal initial state with $\chi_r\approx0$ (left column, corresponding to data as in Fig.~\ref{F4}) and for an initial state with non-negligible linear tail $\chi_r\neq0$ (right column, corresponding to data as in Fig.~\ref{F7}). The evolution in the absence (orange (bright) lines, see Ref.~\cite{langen2013local}) and in the presence (blue (dark) lines) of phonon-phonon collisions is compared for increasing times $t=(1,2,3,4)/(40 v\Lambda^2)$.}
	\label{F11}
\end{figure*}

{\it{Dynamics without collisions}}  --
In the absence of collisions, the time evolution of the relative phase is determined solely by the dephasing of the anomalous modes $m_{r,p}$. Eliminating constant terms, which vanish upon normalization of $C$, the time dependent relative phase is
\begin{widetext}
\begin{align}
\Delta \theta_{r,x,t}=\int_p\frac{e^{-\frac{|p|}{\Lambda}}}{2K|p|}[1-\cos(px)][2n_{r,p,0}+1-2m_{r,p,0}\cos(2\nu|p|t)]\approx\begin{cases} 
      \frac{\pi T_{r}}{2K\nu}x & x < 2\nu t \\
     \frac{\pi T_{r}}{K} t& 2\nu t < x \\
   \end{cases},\label{E49}
\end{align}
\end{widetext}
where $m_{r,p,0}=\frac{T_r}{\nu|p|}-\chi_r|p|$ and $n_{r,p,0}=\frac{T_r}{\nu|p|}+\chi_r|p|-\frac{1}{2}$. The approximation on the right of \eqref{E49} is valid for $x\gg a$, i.e. distances larger than the short-distance cutoff. It shows two characteristic properties: First, there is no functional dependence on $\chi_r$, which modifies only the short distance modes at larger momenta $p>\sqrt{T_r/(\nu\chi_r)}$. The linear tail $\sim \chi_r|p|$ appears only as a global shift of $\Delta\theta_{r,x,t}$, which is not resolved in experiments. Second, the light cone $x=2\nu t$ clearly separates the initial state regime $x>2\nu t$ from the prethermal regime $x<2\nu t$. This feature of the relative phase has been exploited in order to experimentally measure the spreading of prethermal correlations and the value the sound velocity in the quantum wires \cite{geiger2014local,gring2012relaxation,langen2013local}. It also determines the ratio $T_r/K$, which depends both on the initial state via $T_r$ and the Hamiltonian via the Luttinger parameter $K$. In Fig.~\ref{F11}, the time evolution of $C, \Delta\theta_r$ in the absence of collisions (orange (bright) lines) is plotted for two distinct initial states with either $\chi_r=0$ or $\chi_r\neq0$. 

{\it{Dynamics in the presence of collisions}} --
In the presence of collisions energy is redistributed from the absolute to the relative mode via the decay of the interwire coherence $n_{12,p,t}\rightarrow0$. For non-zero initial $\chi_r$, there will be an additional transfer of energy from larger to lower momenta in order to establish a thermal phonon distribution within each wire. This process is monitored by a non-vanishing deviation $\Delta n_{r,p,t}=n_{r,p,t}-n_{r,p,0}$, which is building up in time. Following the discussion of the time evolution of $n_{r,p,t}$ above, $\Delta n_{r,p,t}$ can be decomposed into two distinct regimes. For momenta lower than the crossover momentum $p<p^{(c)}_t$, $\Delta n_{r,p,t}\sim |p|$, while for larger momenta $p>p^{(c)}_t$ it is that $\Delta n_{r,p,t}\sim 1/|p|$. 

On distances $x>2\nu t$, the anomalous contributions $\sim m_{r,p,t}$ have already dephased and the relative phase can be approximated as
\begin{align}
\Delta \theta_{r,x,t}=&\int_p\frac{e^{-\frac{|p|}{\Lambda}}}{2Kp}[1-\cos(px)][2n_{r,p,0}+1+2\Delta n_{r,p,t}]\nonumber\\ \approx&\left\{\begin{array}{cl}\frac{\pi x}{2K\nu}T\sub{f},& \text{ for } p^{(c)}_tx<1\\ \frac{\pi x}{2K\nu}\left(T_r+(T\sub{f}-T_r)p^{(c)}_tx\right),& \text{ for } p^{(c)}_tx>1\end{array}\right. ,\label{E51}
\end{align} 
where $T\sub{f}$ is again the final temperature in the relative mode, determined via Eq.~\eqref{asympT}, and $T_r$ depends on the initial condition.
The crossover distance in Eq.~\eqref{E51} can be estimated with $x^{(c)}_t=1/p^{(c)}_t\approx (T_r v^2/\nu)^{-1/3} t^{2/3}$. For $x^{(c)}_t>a$ larger than the short distance cutoff, it spreads slower than the prethermal distance $x\sub{pre}=2\nu t$ and thus the decomposition of Fig.~\ref{F5} applies. 

Along the same lines, one can determine the relative phase for distances $x>2\nu t$, which yields after some algebra
\begin{align}
\Delta \theta_{r,x,t}\overset{x>2\nu t}{\approx}\frac{\pi t}{K}\left(T_r+2\nu(T\sub{f}-T_r)p^{(c)}_tt\right).
\label{EE51}
\end{align} 
The additional term in this equation compared to the collisionless case in Eq.~\eqref{E49} grows in time $\sim (T\sub{f}-T_r)t^2p_t^{(c)}=(T\sub{f}-T_r)t^{4/3}$ for $p_t^{(c)}\sim t^{-2/3}$, i.e. it is proportional to the difference of initial and final temperature and grows faster than linear in time, witnessing interaction effects. 

As for the collisionless case, Eqs.~\eqref{E51}, \eqref{EE51} do not contain any explicit dependence on $\chi_r$. A non-zero $\chi_r$ in the initial state only enters the final temperature $T\sub{f}$ since the energy stored in the linear tail is redistributed over the whole momentum range. In total, the phase correlation function $C$ can hardly distinguish between a thermal state $\chi_r=0$ where the only increase in the relative temperature results from the equilibration with the absolute mode or a non-thermal state $\chi_r>0$, for which in addition to the equilibration with the absolute mode, energy is redistributed within the relative mode as well. Only significantly large $\chi_r\Lambda\gg T_r/(\nu\Lambda)$ would change this observation, by drastically increasing $T_f$, which would, however, as well prohibit to work in the low energy Luttinger framework.
 Furthermore, the spatial structure of $\Delta\theta_{r,x,t}$ in the thermal ($xp^{(c)}_t<1$) and prethermal regime ($xp^{(c)}_t>1$) are almost indistinguishable within a real time evolution of $\Delta\theta_{r,x,t}$ and $C_{r}$ as shown in Fig.~\ref{F11}. 

From this analysis, we must conclude that the thermalization dynamics of the two wire system can hardly be investigated on the basis of initial states of the form \eqref{E15}-\eqref{E16}. One possible, minor witness for the presence or absence of thermalization, could be the additional, non-linear growth of $\Delta\theta_{r,x,t}$ in time in the regime $x>2\nu t$ as expressed in Eq.~\eqref{EE51}. Despite the $T_f - T_r$ may be substantial in experiments, however, the nonlinear dependence of the relative phase would be hard to resolve in practice, cf. Fig.~\ref{F11}.

\subsubsection{Time evolution of the coherence factor}\label{A6}
The coherence factor $\Psi_c(t)=\rho_0e^{-g(t)}$ as defined in Eqs.~\eqref{phid}, \eqref{phid2} measures the integrated coherence between the two wires. It only depends on the absolute forward time and is expected to decay as time evolves since the coherence between the two uncoupled wires decays due to dephasing and collisions. The decay of the coherence factor has been investigated in previous works, which found different asymptotic behavior, depending on the relevant decay mechanism. In Ref.~\cite{bistritzer2007intrinsic}, a linear time-dependence $g(t)=g_0t$ of the exponent has been predicted by the analysis of the collisionless model, while in Ref.~\cite{burkov2007decoherence}, a stretched exponential $g(t)\sim g_1 t^{2/3}$ has been found by assuming energy exchange via collisions of the relative mode phonons from a thermal bath. Here, we will consider the most general case, which combines dephasing as well as phonon-phonon collisions as possible decoherence mechanisms and whose results do not rely on a particular initial state. We begin the discussion with a nearly thermal initial state, $\chi_r=0$, and then extend the scenario to initial states with $\chi_r\neq0$.

{\it{Nearly thermal initial state}} -- For initial states with $\chi_r=0$, the diagonal phonon density $n_{11,p,0}=\frac{T_r+T_a}{2\nu|p|}$ and the retarded self-energy $\sigma^R_q$ in Eq.~\eqref{E36} can be solved analytically. The self-energy takes the thermal form \cite{andreev1980hydrodynamics,punk2006collective,buchhold2015kinetic,burkov2007decoherence} and reads as
\begin{equation}
\sigma^R_q=0.789v\sqrt{\frac{\pi (T_r+T_a)}{\nu}}|q|^{\frac{3}{2}}=\gamma |q|^{\frac{3}{2}},\label{selfen}
\end{equation}
where we have absorbed all momentum independent terms in the prefactor $\gamma$. The dynamics of the anomalous and the interwire phonon mode $m_{11,q,t}, n_{12,q,t}$ is not constrained by any conservation laws and they decay in this case as $m_{11,q,t}=\exp(-\gamma|q|^{3/2}t)m_{11,q,t=0}$ and $n_{12,q,t}=\exp(-\gamma|q|^{3/2}t)n_{12,q,t=0}$. The diagonal phonon mode $n_{11,q,t}$ on the other hand is thermally occupied and does not evolve in time.

\begin{figure}
	\includegraphics[width=\linewidth]{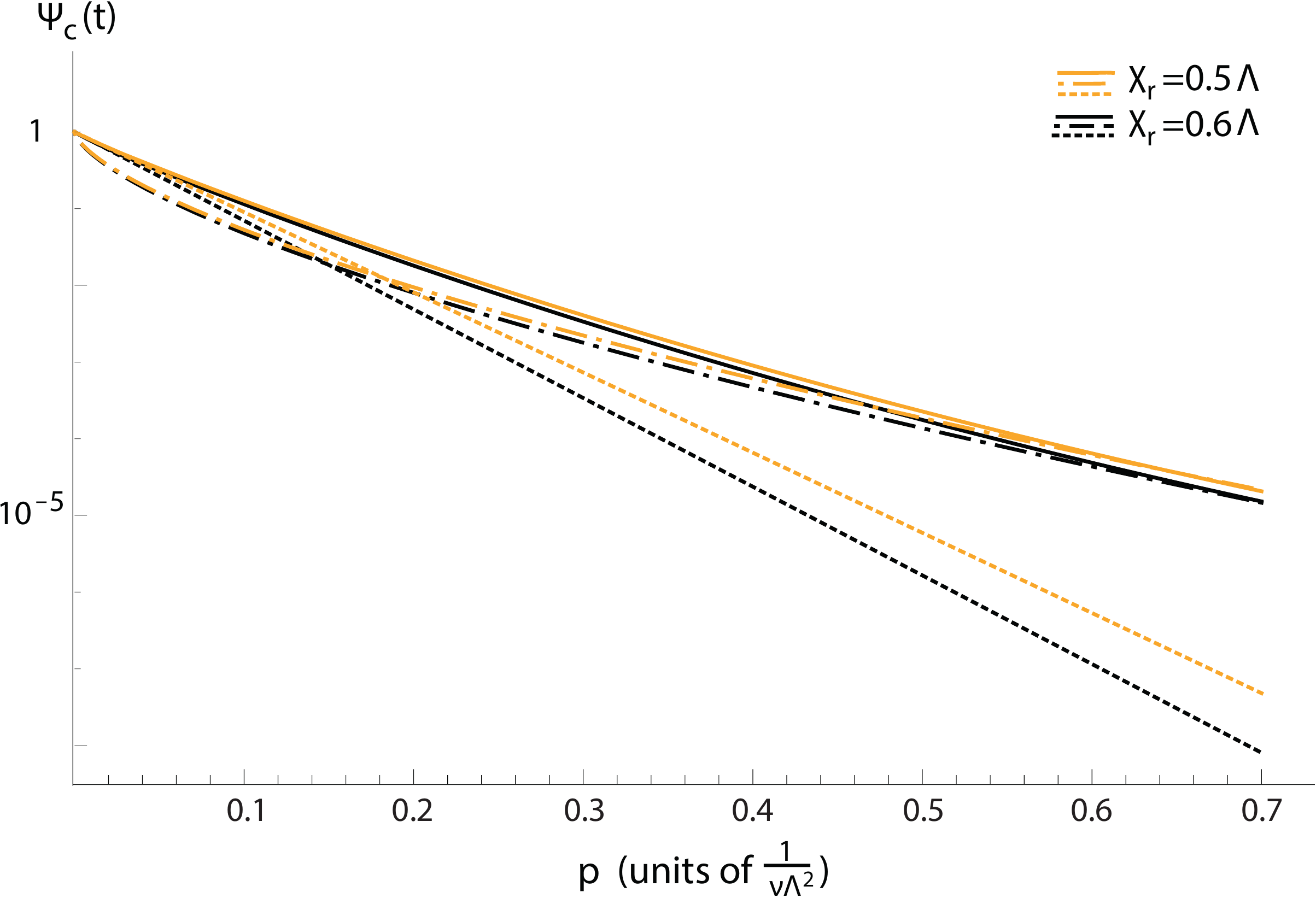}
	\caption{Time evolution of the coherence factor $\Psi_c(t)$ from the kinetic equations (bold line) compared to an exponential decay $\sim \exp(-t/t_0)$ (dashed line), as predicted from pure dephasing, and a stretched exponential $\sim \exp(-(t/t_0)^{2/3})$ (dash-dotted line), as predicted from phonon collisions, see Eqs.~\eqref{E44}, \eqref{lalilu}. The colors (bright and dark) represent two different initial states with $\chi_r=0.5\Lambda^{-1}$ and $\chi_r=0.6\Lambda^{-1}$.}
	\label{F8} 
\end{figure}

The exponent of the coherence factor in this case can be computed analytically
\begin{align}
g(t)=& \int_p\frac{e^{-\frac{|p|}{\Lambda}}}{8K|p|}(2n_{r,p,t}+1-2\cos(2\nu|p|t)m_{r,p,t})\nonumber\\ 
\approx&\int_p \frac{e^{-\frac{|p|}{\Lambda}}}{4K\nu p^2}\left[T\sub{f}+e^{-\gamma|p|^{\frac{3}{2}}t}\left(T_r-T\sub{f}-\cos(2\nu|p|t)T_r\right) \right]\nonumber\\
=&\frac{T\sub{f}}{2K\nu}(\gamma t)^{\frac{2}{3}}f_{y,T_r/T\sub{f}}.
\label{E44}
\end{align}
The shape of the dimensionless function $f_{y,T_r/T\sub{f}}$ is determined by the value of the dimensionless parameter $y=t\nu^3\gamma^{-2}$. It has the general asymptotic behavior
\begin{equation}
f_{y,\tau}=\left\{\begin{array}{cl}\Gamma(\frac{1}{3}),& \text{ for } y\ll1 \\ \pi y^{\frac{1}{3}}\tau, & \text{ for } y\gg 1\label{lalilu}
\end{array}\right. .
\end{equation}
Thus it is dominated by thermal collisions for small times $t<\frac{\gamma^2}{\nu^3}=0.789^2\pi T\sub{f}$ and dominated by dephasing for larger times $t>0.789^2\pi T\sub{f}$ depending on $T_f$ and therefore, via Eq.~\eqref{asympT}, implicitly on the initial state. This behavior is counterintuitive since one expects quantum effects to influence the short time rather than the long time behavior. 
It can, however, be explained by a proper decomposition of the integral into high- and low-momentum modes and is supported by the numerical evaluation of the coherence factor in Fig.~\ref{F8}. The latter starts with a $\log\psi_c \sim t^{2/3}$ curvature for small times and becomes less curved at large times in a logarithmic representation. 

A non-linear initial decay of the coherence factor could be observed in experiments, as suggested in previous works \cite{burkov2007decoherence}, where the absolute modes $n_{a,p,t}$ have been treated as a bath for the relative modes, which does not relax at all. For short times, where the distribution of $n_{a,p,t}$ has not yet been modified by collisions, this is a reasonable assumption and thus coincides with our short time prediction. Compared to the behavior of $g(t)$ in the non-interacting case $g(t)=\frac{\pi T_r t}{2K}$ for all $t>0$ is a signature of collisions and goes beyond the quadratic Luttinger model.

{\it{Non-thermal initial state}} -- For initial states with $\chi_r\neq0$, the major difference compared to the previous case of $\chi_r=0$ is the redistribution of energy within the diagonal mode, which leads to a non-trivial ${\Delta n_{11,p,t}=n_{11,p,t}-n_{11,p,t=0}\neq0}$. Since the exponential decay of the anomalous and interwire modes is dominated by the Rayleigh-Jeans divergence in the diagonal density, it is not expected to be crucially modified. Performing the decomposition $n_{11,p,t}=n_{11,p,t=0}+\Delta n_{11,p,t}$ and inserting this into the definition of $g(t)$, Eq.~\eqref{phid2}, we find $g(t)=g_0(t)+\Delta g(t)$ with
\begin{align}
\Delta g(t) = &\int_p\frac{e^{-\frac{|p|}{\Lambda}}}{4K|p|}\Delta n_{11,p,t}\approx\frac{3T\sub{f}}{4K \nu p^{(c)}_t}\nonumber \\ \approx&\frac{3}{4K}\left(\frac{T\sub{f}^4v^2}{\nu^4}\right)^{\frac{1}{3}}t^{\frac{2}{3}}\label{E45}.
\end{align}

Due to energy redistribution within a single wire, the coherence factor thus shows again a stretched exponential decay with an exponent $ \sim t^{2/3}$ even at the largest times. This stretched exponential decay of the coherence factor at the shortest and the longest time scales represents a clear sign of phonon-phonon collisions, either leading to a decay of the inter-wire correlations $\sim n_{12,p,t}$ or the redistribution of energy in the intra-wire density $n_{11,p,t}$. In Fig.~\ref{F8}, the time evolution of the numerically simulated coherence factor is shown to significantly deviate from a linear exponential decay at all times. For short and intermediate times, it is faster than the analytically predicted $\exp(-(t/t_0)^{2/3})$ but it approaches this behavior in the limit $t\rightarrow\infty$, as predicted from the simplified analysis above.

\begin{figure*}
  \includegraphics[width=\linewidth]{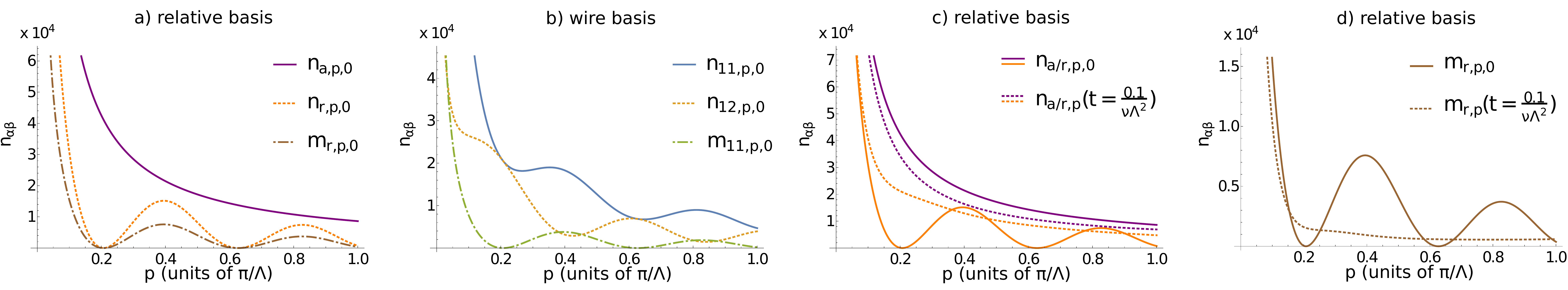}
\caption{Distribution for a non-thermal initial state with parameters as in Table~\ref{Tabel1} and an oscillation length $\xi=7.5/\Lambda$: a) The initial relative and absolute occupations and b) the intra- and inter-wire occupations. c+d) Time evolution of the relative and absolute density occupations in the presence of phonon-phonon collisions. The collisions lead to an adjustment of $n_{r,p,t}$ and $n_{a,p,t}$ as time evolves, which is accompanied with the temporal fade out of the scale $\xi$.}
\label{F12}
\end{figure*}

\section{Splitting to a non-thermal initial state}\label{A8}
In order to observe the thermalization dynamics in the two-wire system much more clearly than in the previous section, it is most promising to design the splitting protocol in such a way that the initial state is no longer close to an effective thermal state but rather shows a clear structure of an out-of-equilibrium state in the normal, relative mode density $n_{r,q,t=0}$. Such a state will then relax very differently depending on whether phonon-phonon collisions are present or absent, making their influence on the dynamics clearly observable. Here, we are inspired by the latest experiments in Vienna, where the initial state has been prepared in such a way, that the occupations of the quantized phonon modes with odd momentum index were much smaller, than the occupations of the even modes. The latter could be described by an effective thermal distribution \cite{langen2015experimental}. Within the short-time evolution probed in the experiment, the two-wire system entered the prethermal regime and relaxed towards a non-thermal, generalized Gibbs ensemble (GGE) \citep{jaynes1957information,rigol2007relaxation,polkovnikov2011colloquium,kollar2011generalized}. Below, we explore the complete time evolution of such an initial state in the presence of phonon-phonon collisions and discuss in which way the thermalization dynamics can be made experimentally accessible by tuning the initial state of the system. 

\subsection{Initial state and relaxation of the phonon densities}
The above mentioned, experimentally observed initial state was realized in a wire of length $L$ for which the low momentum modes were approximately quantized to integer multiples of $\pi/L$. By modifying the splitting protocol, only the even modes were considerably populated, while the odd momentum modes remained empty \cite{langen2015experimental}.
Here we generalize this state by going to the continuum of momentum modes and introducing a flexible length scale $\xi$ for the oscillations of the phonon occupations. The phonon occupation shall be maximal for momenta $p=2\alpha\pi/\xi $ and minimal for $p =(2\alpha+1)\pi/\xi$, where $n$ is an integer. Thus, for $\xi=L/2$ the experimentally observed state is recovered. Motivated by this observation, we consider an initial phonon distribution of the form (see Fig. \ref{F12}):
 \begin{align}
n_{r,p,t=0}&=\frac{T_r}{\nu |p|}\cos^{2}(p \xi)-\frac{1}{2},\label{E53}\\
n_{a,p,t=0}&=n_{B}(\nu|p|/T_a),\label{E54}\\
m_{r,p,t=0}&=\frac{T_r}{\nu |p|}\cos^{2}(p\xi).\label{E55}
\end{align}
Here, $T_r$ is the effective temperature of the envelope function $\sim \frac{T_r}{\nu|p|}$, as in the previous section. We will consider, however, a more general oscillation length scale $0\le \xi\le L/2$, which interpolates continuously between the GGE state in Ref.~\cite{langen2015experimental} for $\xi=L/2$ and the previously discussed, nearly thermal state \cite{hofferberth2007non,kitagawa2011dynamics,geiger2014local,gring2012relaxation,langen2015experimental}, obtained for $\xi=0$. 

An initial relative density of the form \eqref{E53} with $\xi=L/2$ has been inferred from the experimental data in Ref.~\cite{langen2015experimental} after splitting one wire within a single, relatively fast splitting operation. The corresponding initial anomalous phonon density $m_{r,p,t}$ is much more difficult to detect in the experiment. It must, however, be of the form \eqref{E55} in order to guarantee a vanishing initial relative phase difference $\Delta\theta_{r,p,t}=0$ between the two wires. 

The nearly thermal relative density \eqref{E15} with $\xi=0$, on the other hand, has been observed experimentally after performing a two-staged splitting protocol, which consists of a slow initial and a much faster final splitting operation \cite{langen2015experimental}. It is therefore reasonable to assume that a general $\xi$ between these two limiting cases can be obtained by continuously modifying the corresponding splitting procedure. 

In the absence of phonon collisions, the dynamics is governed by the dephasing of $m_{r,p,t}$, which drives the system into the prethermal regime. The corresponding steady state is described by a GGE, respecting the initial structure of the normal relative density $n_{r,p,t}$, which in the absence of collisions is an integral of motion. In the presence of phonon collisions, on the other hand, the relative and the absolute phonon density both relax towards a thermal distribution with a uniform temperature $T\sub{f}$. 

As in the previous section, the temperature $T\sub{f}$ can be determined from the initial state by exploiting the conservation of the total kinetic energy. Analogous to Eq.~\eqref{asympT} this yields the final temperature
\begin{equation}
T\sub{f}=\frac{T_r+T_a}{2}-\frac{T_r(\xi\Lambda)^2}{1+4(\xi\Lambda)^2}.\label{tfeq}
\end{equation}
It reduces to the previously discussed expression $T\sub{f}=(T_r+T_a)/2$ in the limit of vanishing oscillation scale and to $T\sub{f}=(T_r/2+T_a)/2$ in the limit $\xi=L\gg\Lambda$. The time evolution of $n_{r,p,t}, n_{a,p,t}$ and the corresponding adjustment of both distributions is shown in Fig.~\ref{F12}.

\begin{figure}
\centering
  \includegraphics[width=\linewidth]{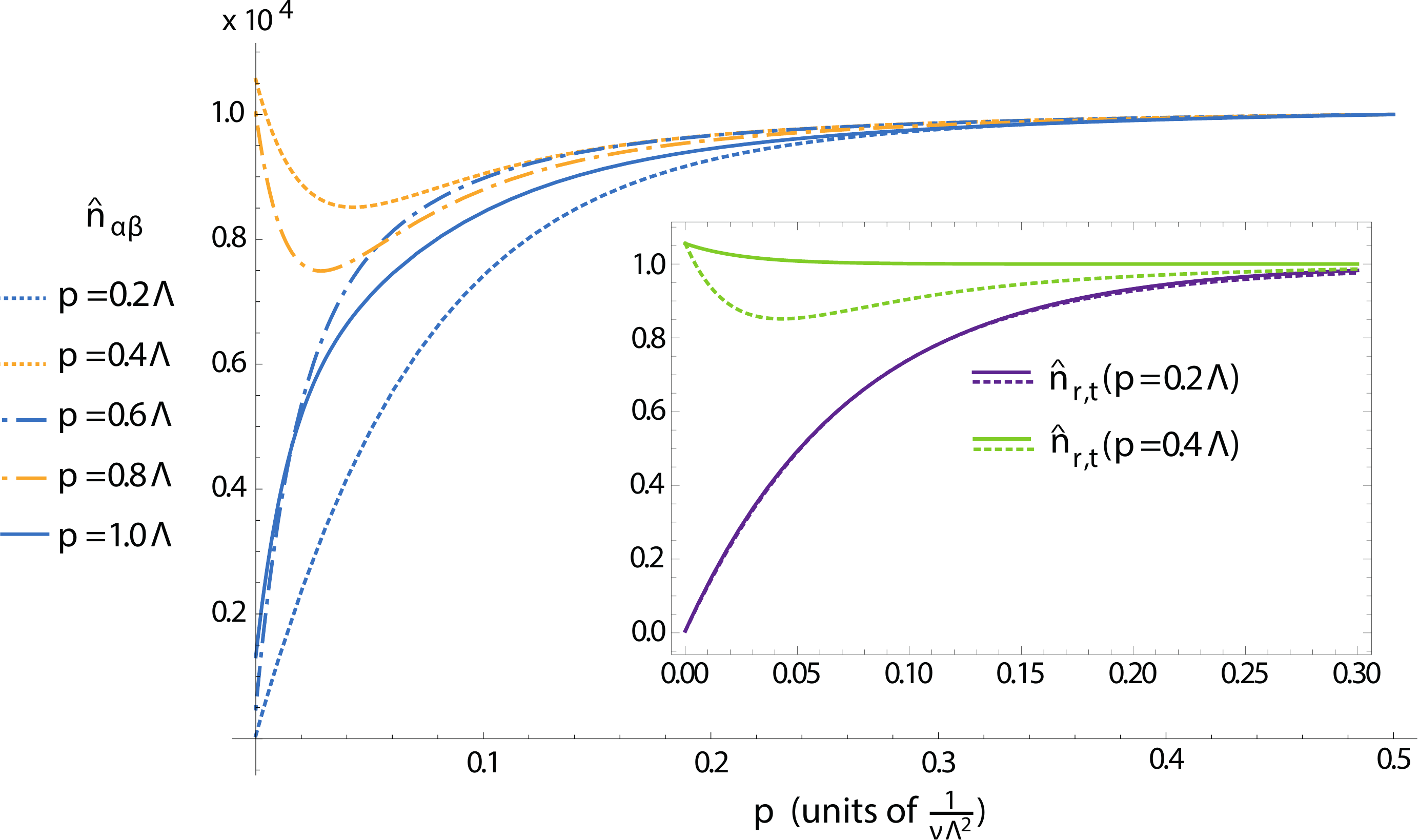}
\caption{Long-time relaxation of the phonon densities $n_{r,p,t}$ for momenta $p=\frac{l\pi}{2\xi}\approx 0.2 l\Lambda$ for $\xi=7.5/\Lambda$ and $l=1,2,...5$, as in Fig.~\ref{F12} ($l=2,4$ correspond to even modes in the experiment \cite{langen2015experimental}). For better visualization, the phonon densities are normalized to their asymptotic, thermal value $\hat{n}_{r,t}(p)=n_{r,p,t}\nu |p|/T_{\text{f}}$. The relaxation can be described in terms of a two-stage process. In the first stage, the relative modes try to approach one common temperature $\sim T_r<T_{\text{f}}$ in order to remove the oscillating structure in $n_{r,p,t}$, i.e. energy is redistributed within the relative modes. This can be achieved already by local scattering processes with momenta $\sim \xi^{-1}$. In the second, slower stage, the temperature between the relative and the absolute mode are adjusted via energy transfer from the absolute to the relative mode. The two stages are most prominently visible in the evolution for the momenta $p=0.4, 0.8 \Lambda$, which first decrease below their asymptotic value and then, together with the other modes, increase again towards $T_{\text{f}}$. The inset shows comparison of a bare exponential decay (bold lines) corresponding to a single stage relaxation process with the numerical data (dotted lines) for $p=0.2\Lambda$ (lower graphs) and $p=0.4\Lambda$ (upper graphs). For the latter, the deviation is very pronounced. 
}
\label{FExtra1}
\end{figure}

One way of experimentally tracing the thermalization dynamics is to detect the time evolution of the phonon densities. The latter can be inferred from the experimental data in the limit of $m_{r,p,t}=0$, i.e. after the complete dephasing of the anomalous densities, when prethermalization has been reached. In this limit, the relative phase is determined by the normal distribution alone. This has been demonstrated successfully in Ref.~\cite{langen2015experimental}. For an initial state considered here, the time evolution of $n_{r,p,t}$ is most pronounced at the local extremums of the initial phonon density at momenta $p=\frac{l\pi}{2\xi}$, with integer $l$. The corresponding evolution is shown in Fig.~\ref{FExtra1} for parameters as in Fig.~\ref{F12}. In the inset, we compare this evolution with the analytical prediction 
\begin{equation}
n_{r,p,t}=n_{r,p,t=0}+(n_{r,p,t=\infty}-n_{r,p,t=0})e^{-\sigma^R_pt}
\end{equation}
for a self-energy $\sigma^R_p=0.789v\sqrt{2\pi T_{\text{f}}/\nu}$ (cf. Eqs.~\eqref{selfen}, \eqref{tfeq}) which we approximate to be time-independent and uniform.

This comparison demonstrates that the thermalization of the phonon modes for this particular initial state is a two-stage process. This is pronouncedly visible in the maximally occupied modes ($p=\frac{2l\pi}{2\xi}$), corresponding to the even modes in the experiment \cite{langen2015experimental}. In the first stage, the oscillatory behavior of $n_{r,p,t}$ is smoothened via local collisions of momenta $\sim \xi^{-1}$. This corresponds to energy exchange between the relative modes. In the second stage, global thermalization is achieved by energy redistribution between the relative and the absolute mode, see Fig.~\ref{FExtra1}. 

\subsection{Relaxation of the phase correlation function}
The oscillatory behavior of the phonon densities introduces an additional length scale, the oscillation length $\xi$, which will be visible in experimental observables that depend on $n_{r,q,t}$, such as the phase correlation function $C(x,t)$ discussed above. The thermalization dynamics due to collisions will, however, lead to a uniform, non-oscillating temperature and erase the effect of $\xi$ in the asymptotic evolution. In the following, we discuss in which way this behavior can be monitored in terms of the phase correlation function. 
\begin{figure*}
\centering
  \includegraphics[width=\linewidth]{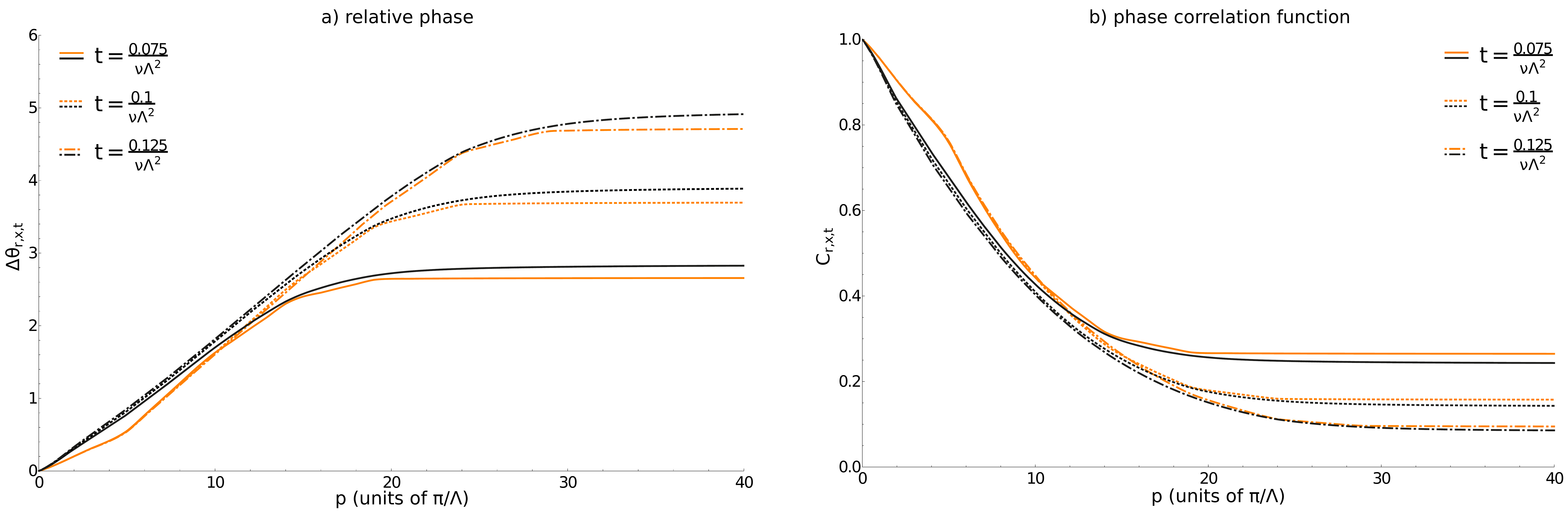}
\caption{Initial parameters as in Table~\ref{Tabel1} and an oscillation length $\xi=7.5\Lambda^{-1}$: a) Comparison of the evolution of the relative phase $\Delta\theta_{r,x,t}$ in the presence (black (dark) lines) and absence (orange (bright) lines) of phonon collisions. b) The same comparison for the coherence factor $C(x,t)$. The difference between a thermal and a non-thermalized state becomes quite significant at distances on the order of twice the oscillation length, $x=2\xi$.}
\label{F13}
\end{figure*}

{\it Relaxation in the absence of phonon collisions} -- 
In the absence of collisions, the dynamics of $\theta_{r,x,t}$ is completely governed by the dephasing of the anomalous contributions and the relative phase is determined by the middle part of Eq.~\eqref{E49} and the initial densities \eqref{E53}-\eqref{E55}, leading to
\begin{equation}\label{relPh}
\Delta\theta_{r,x,t}=\int_p \frac{T_re^{-\frac{|p|}{\Lambda}}}{K\nu p^2}[1-\cos(px)][1-\cos(2\nu|p|t)]\cos^2(p\xi).
\end{equation}
This equation can be solved analytically but contains quite a number of different temporal and spatial regimes. Since we are generally interested in the thermalization dynamics at distances $x\sim \xi$, we assume $\nu t\gg \xi$ in the following. In this limit, we find
\begin{equation}\label{relPh2}
\Delta\theta_{r,x,t}=\frac{\pi T_r}{4K\nu}\cdot\left\{\begin{array}{cl}x, & \text{ for }x<2\xi\\ 2x-2\xi, & \text{ for }2\xi<x<2\nu t \\ 4\nu t-2\xi, & \text{ for } 2\nu t<x\end{array}\right. 
\end{equation}
and recognize a change in the slope of $\Delta\theta_{r,x,t}$ at distances $x=2\xi$. For smaller distances $x<2\xi$, the slope is only half as steep as for larger distances $x>2\xi$. 

This leads to a clearly observable cusp in the spatial dependence of $\Delta\theta_{r,x,t}$ as well as in $C(x,t)$ at $x=2\xi$, as one can see from Fig.~\ref{F13} (orange (bright) lines). This cusp is observable in the prethermal regime $t\sub{ther}>t>t\sub{preth}$, for which $m_{r,p,t}$ is dephased but $n_{r,p,t}$ remains still pinned to its initial value $n_{r,p,t=0}$.
One does, however, not expect such behavior in the thermal regime for times $t>t\sub{therm}$ and thus to serve as an indicator of thermalization on distances $x\sim 2\xi$.

{\it Relaxation in the presence of collisions} -- In the presence of collisions, the normal relative density $n_{r,p,t}$ is no longer prevented from relaxation and acquires the shape of a thermal Rayleigh-Jeans distribution, $n_{r,p,t\rightarrow\infty}=T\sub{f}/(\nu|p|)$. The thermalized momentum regime, i.e. the momentum regime $p>p^{(c)}_t$, for which the relative density is well described by this distribution, is established at larger momenta and spreads to lower momenta as time evolves. As soon as the thermalized regime reaches the inverse oscillation scale, $2p^{(c)}_t=\xi^{-1}$, the effect of collisions becomes visible on the length scale of the oscillations. Consequently, the relative phase on smaller distances $x<2\xi$ is thermal $\Delta\theta_{r,x,t}=\frac{\pi T\sub{f}x}{2K\nu}$. In Fig.~\ref{F13} a), the time evolved relative phase $\Delta\theta_{r,x,t}$ in the presence of phonon collisions is compared to the $\Delta\theta_{r,x,t}$ in the absence of collisions and the difference is clearly visible at $x=2\xi=15\Lambda^{-1}$. 

In the limit of $t\rightarrow\infty$, one can evaluate the ratio of the thermal coherence factor $C^{(\text{therm})}$ over the coherence factor in the absence of collisions $C^{(\text{preth})}$:
\begin{equation}
c_\xi=\frac{C^{(\text{therm})}(2\xi,t)}{C^{(\text{preth})}(2\xi,t)}=e^{\frac{\pi \xi}{4K\nu}\left(T_r-2T\sub{f}\right)}=e^{\frac{\pi \xi}{4K\nu}\left(\frac{2T_r(\xi\Lambda)^2}{1+4(\xi\Lambda)^2}-T_a\right)}.
\end{equation}
For the choice of a moderate oscillation length $\xi\approx 7.5\Lambda^{-1}$ and the parameters from Tab.~\ref{Tabel1}, $c_\xi\approx0.83$, which is clearly within the experimental resolution of the coherence factor.

The typical time scale at which the coherence factor of the thermalizing system starts to deviate from the prethermal coherence factor at $x=2\xi$ is given by the collision time at momentum $p=\xi$,  $t_{\text{coll}}=\big(\nu/T_r v^2\big)^{1/2}\xi^{\frac{3}{2}}$. For a set of initial states with different oscillation lengths $\xi$ this would also allow the measurement of the exponent $\alpha=3/2$ for the self-energy, validating the picture of resonantly interacting phonons in one dimension.

In Fig.~\ref{F13} b), the evolution of the coherence factor for a thermalizing system with an oscillating initial distribution is compared to its evolution in the absence of collisions. For the chosen evolution times $t$, the evolution at the distance $x=2\xi$ has taken place and both graphs are static but well separable. This demonstrates the clear difference between the prethermal state of the system (corresponding to an evolution in the absence of collisions) and the asymptotic thermal state for this class of initial states and thus $c_\xi$ can serve as a tool for the experimental observation of thermalization in the two wire system. In contrast to the evolution starting from a nearly thermal state, where asymptotic thermalization is hardly observable in the long distance behavior of the coherence factor (cf. Fig.~\ref{F8}), the evolution of an oscillating initial density features a clear, short-distance ($x\approx 2\xi$) signature of thermalization. Given the experimental realizability of such out-of-equilibrium states, which is motivated here based on previous experiments, they may serve as good candidates for the observation of thermalization in atomic interferometers.\\

\section{Conclusion}\label{Conc}
In this article, we have investigated the relaxation of two initially correlated quantum wires after an initial split in the presence of intra-wire phonon collisions. We put an emphasis on comparing this dynamics to the previously studied collisionless Luttinger liquid theory and on the analysis of the experimental observability of the asymptotic thermalization in the interacting model. While the effect of collisions is clearly visible in terms of the temporal decay of inter-wire correlations $\langle a_{1,x,t}^{\dagger}a_{2,x,t}\rangle\overset{t\rightarrow\infty}{\rightarrow} 0$ and the redistribution of energy between the intra-wire modes, it is less pronounced in typical experimental observables like the coherence factor $\Psi_c(t)$ or the relative phase correlation function $C(x,t)$. This sheds light on the experimental observation that, even for relatively large times, a deviation of the dynamics from the quadratic Luttinger theory prediction is hardly observable. The latter relies on the fact that for the common splitting procedure the resulting initial state is very close to a thermal state, which features a discrepancy in the temperatures between the relative and the absolute mode. 
We demonstrate, however, that indications for thermalization can become more pronounced if the initial state is prepared farther away from a thermal state, for instance by modifying the splitting procedure. By tuning the initial state such that it becomes clearly non-thermal, the effect of collisions and the dynamics beyond single-particle dephasing can be clearly observed in experiments. This lays out a potential route for further experiments in order to reveal the asymptotic thermalization dynamics of nearly integrable models.

\emph{Acknowledgements} -- We thank B. Rauer for fruitful discussions. S. H. was supported by the German Excellence Initiative via the Nanosystems Initiative Munich (NIM). M. B. and S. D. acknowledge support by the German Research Foundation (DFG) through the Institutional Strategy of the University of Cologne within the German Excellence Initiative (ZUK 81), and S. D. support by the DFG within the CRC 1238 (project C04) and the European Research Council via ERC Grant Agreement n. 647434 (DOQS). M. B. acknowledges support from the Alexander von Humboldt foundation. J. S. acknowledges support by the European Research Council, ERC-AdG {\it QuantumRelax} (320975).

\bibliography{InteractingLL}

\appendix
\setcounter{equation}{0}
\setcounter{section}{0}
\setcounter{figure}{0}
\setcounter{table}{0}
\setcounter{page}{1}

\renewcommand{\theequation}{A\arabic{equation}}
\renewcommand{\thefigure}{A\arabic{figure}}
\renewcommand{\bibnumfmt}[1]{[S#1]}
\renewcommand{\citenumfont}[1]{S#1}

\section{Kinetic Equation from the Keldysh Formalism}\label{AS1}
We give a short overview on the central concepts and formulas of the Keldysh field integral formalism that are used in the main work. The first section is devoted to the description of the Green's function in Nambu basis for the coupled two-wire system. Then we go on and introduce the Wigner transformation and finally discuss the derivation of the kinetic equations.
\subsection{Nambu space representation}
In the presence of anomalous modes, the Green's functions and correlations are conveniently expressed in terms of Nambu spinors of complex bosonic fields in Keldysh space \citep{kamenev2011field} 
\begin{align}
\ann{A}{\alpha,p,t}=&(\ann{A}{cl,\alpha,p,t},\ann{A}{q,\alpha,p,t})\label{ES1}\\
=&(\ann{a}{cl,\alpha,p,t},\cref{a}{cl,\alpha,-p,t},\ann{a}{q,\alpha,p,t},\cref{a}{q,\alpha,-p,t})^T,\nonumber
\end{align}
where $(cl,q)$ label classical and quantum fields and $\alpha$ is either the wire index $\alpha=1,2$ or labels the relative and absolute modes $\alpha=a,r$.
The total Nambu Green's function in the wire basis is defined as
\begin{align}
iG_{p,t,t'}=
\langle
\begin{pmatrix}
\ann{A}{1,p,t} \\
\ann{A}{2,p,t}
\end{pmatrix}
\begin{pmatrix}
\cre{A}{1,p,t'} ,&& \cre{A}{2,p,t'}
\end{pmatrix}
\rangle.\label{ES2}
\end{align} 
It consists of $4$ sectors $\alpha,\beta$ whereas each sector consists of advanced/retarded and Keldysh (correlation) Green's function
\begin{align}
 iG_{\alpha\beta,p,t,t'}=&\langle \ann{A}{\alpha,p,t} \cre{A}{\beta,p,t'} \rangle\label{ES3}\\
=&
\begin{pmatrix}
G^{K}_{\alpha\beta,p,t,t'} && G^{R}_{\alpha\beta,p,t,t'}\\
G^{A}_{\alpha\beta,p,t,t'} && 0
\end{pmatrix},\nonumber
\end{align}
where, again, $\alpha,\beta$ are wire indexes. 
The retarded Green's function in operator representation reads ($[ , ]$ the commutator)
\begin{align} 
iG^{R}_{\alpha\beta,p,t,t'}
&=\Theta(t-t')
\begin{pmatrix}
\langle [\ann{a}{\alpha,p,t},\cre{a}{\beta,p,t'}]  \rangle & \langle [\ann{a}{\alpha,p,t},\ann{a}{\beta,-p,t'} ]  \rangle \\
\langle [\cre{a}{\beta,p,t'},\cre{a}{\alpha,-p,t} ]  \rangle & \langle [\cre{a}{\beta,-p,t'},\ann{a}{\alpha,-p,t} ]  \rangle
\end{pmatrix}\nonumber\\
&=i\begin{pmatrix}
d^{R}_{\alpha\beta,p,t,t'} & o^{R}_{\alpha\beta,p,t,t'} \\
o^{A}_{\alpha\beta,-p,t,t'} & d^{A}_{\alpha\beta,-p,t,t'}
\end{pmatrix}.\label{ES4}
\end{align}
Advanced and retarded propagators are hermitian conjugate to each other $G^{A}_{\alpha\beta,p,t,t'}=[G^{R}_{\alpha\beta,p,t,t'}]^{\dagger}$. Since the Hamiltonian does not couple the two different wires, annihilation and creation operators of different wires commute at all times, i.e. $[\ann{a}{\alpha,p},\cre{a}{\beta,p'}]=\delta_{\alpha,\beta}\delta_{p,p'}$ and thus the retarded Green's function is diagonal in the wire index.\\
The Keldysh Green's function in the wire basis has the operator representation ($\{,\}$the anti-commutator)
\begin{align} 
iG^{K}_{\alpha\beta,p,t,t'}&=
\begin{pmatrix}
\langle \{\ann{a}{\alpha,p,t},\cre{a}{\beta,p,t'}\}  \rangle & \langle \{\ann{a}{\alpha,p,t},\ann{a}{\beta,p,t'} \}  \rangle \\
\langle \{\cre{a}{\alpha,p,t},\cre{a}{\beta,p,t'}\} \rangle & \langle \{\ann{a}{\alpha,p,t},\cre{a}{\beta,p,t'} \} \rangle
\end{pmatrix}\nonumber\\
&=i\begin{pmatrix}
d^{K}_{\alpha\beta,p,t,t'} & o^{K}_{\alpha\beta,p,t,t'} \\
o^{K}_{\alpha\beta,-p,t,t'} & d^{K}_{\alpha\beta,-p,t,t'}
\end{pmatrix}.\label{ES5}
\end{align}
It is anti-hermitian $G^{K}_{\alpha\beta,p,t,t'}=-[G^{K}_{\alpha\beta,p,t,t'}]^{\dagger}$, which allows us to parametrize it in terms of the hermitian distribution function $F_{p,t,t'}$ and the retarded/ advanced Green's function, as well as the matrix $\Sigma_z=\sigma_z\otimes\mathds{1}$ via
\begin{align}
G_{p,t,t'}^{K}=\left(G^{R} \circ \Sigma_{z}\circ F-F\circ \Sigma_{z} \circ G^{A}\right)_{p,t,t'}.\label{ES6}
\end{align}
The matrix $\Sigma_z$ thereby preserves the symplectic structure of the bosonic Nambu space and $\circ$ represents a convolution in time and a product in momentum space.

\subsection{Wigner approximation and Dyson equation}
The two-point functions in this work are represented in Wigner coordinates in time. It is a convenient representation for non-time-translational invariant systems and for two different times $t_1, t_2$ introduces a forward (or absolute) time $t=(t_1+t_2)/2$ and a relative time $\delta_t=t_1-t_2$ \cite{kamenev2011field}. The Wigner transform $\mathcal{WT}$ is the Fourier transform with respect to the relative time argument, we denote the corresponding Green's function as
\begin{align}
G_{p,t,\omega}=\mathcal{WT}\{G_{p,t_1,t_2}\}=\int_{\delta_t} e^{-i \omega \delta_t}G_{p,\ t+\frac{\delta_t}{2},\ t-\frac{\delta_t}{2}}.\label{ES8}
\end{align}

In order to determine the time evolution of the distribution function $F$, which contains the phonon densities, one makes use of the Dyson equation
\begin{align}
G^{-1}=&P-\Sigma,\label{ES13}
\end{align}
where $G$ is the full propagator and $P$ is the inverse, bare propagator and $\Sigma$ is the self-energy. 
It has the structure
\begin{align}
\Sigma_{\alpha\beta,p,\omega,t}=\begin{pmatrix}
0 & \Sigma^{A}_{\alpha\beta,p,\omega,t}\\
\Sigma^{R}_{\alpha\beta,p,\omega,t} & \Sigma^{K}_{\alpha\beta,p,\omega,t}
\end{pmatrix}.\label{ES14}
\end{align}
As discussed in section \ref{A2}, the resonant interactions couple only phonons that travel in the same direction and thus no pairs $p, -p$, which excludes anomalous retarded self-energies. Furthermore, the Hamiltonian contains no inter-wire coupling, such that the retarded self-energy must be also diagonal in the wire basis (please note that both of these statements do not hold for the Keldysh self-energy). It thus reads as
\begin{align}
\Sigma^{R}_{\alpha\beta,p,\omega,t}&=-i\delta_{\alpha,\beta}\begin{pmatrix}
\sigma_{\alpha\beta,p,\omega,t}^{R} & 0\\
 0 & \sigma_{\alpha\beta,-p,-\omega,t}^{A} \end{pmatrix}.\label{ES15}
\end{align} 
The anti-hermitian Keldysh self-energy is 
\begin{align}
\Sigma^{K}_{\alpha\beta,p,\omega,t}&=-i2
\begin{pmatrix}
\sigma_{\alpha\beta,p,\omega,t}^{K} & \Phi_{\alpha\beta,p,\omega,t}^{K}\\
(\Phi_{\alpha\beta,p,\omega,t}^{K})^{\ast} & \sigma_{\alpha\beta,p,\omega,t}^{K}
\end{pmatrix}.\label{ES16}
\end{align}
\begin{figure}
	\includegraphics[width=0.7\linewidth]{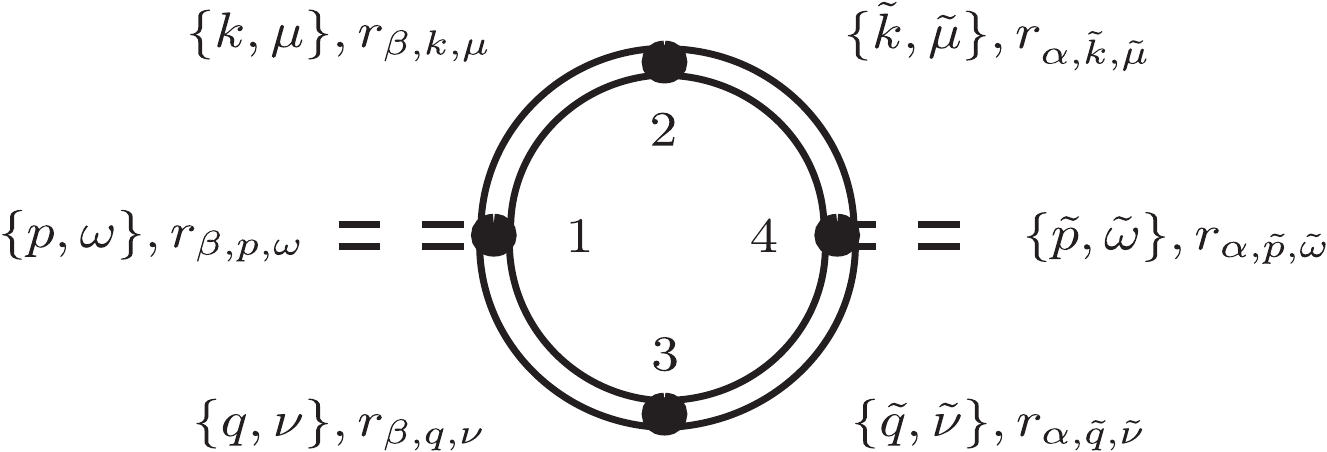}
	\caption{One-loop diagram for the self-energy $\Sigma$. Each line has momentum and frequency $\{k,\omega\}$ and a direction $r_{\alpha,k,\mu}=\pm 1$ for outgoing, ingoing lines, which depends also on the wire index $\alpha$. 
	 Momentum conservation, energy conservation and the resonance condition have to be applied at each connecting element $\{1,2,3,4\}$, which requires that normal (anomalous) self-energies are generated exclusively by normal (anomalous) Green's functions.}
	\label{FS1}
\end{figure}

The kinetic equation \eqref{E24} is now determined by substituting equation \eqref{ES13} into \eqref{ES6} and rearranging it to
\begin{align}
\left(P^{R} \circ \Sigma_{z}\circ F-F\circ \Sigma_{z} \circ P^{A}\right)_{p,\omega,t}=I^{\text{coll}}_{p,\omega,t}. \label{ES17}
\end{align}
The collision integral on the right side is defined as
\begin{align}
I^{\text{coll}}_{p,\omega,t}=\Sigma^{K}_{p,\omega,t}-\left(\Sigma^{R} \circ \Sigma_{z}\circ F-F\circ\Sigma_{z} \circ \Sigma^{A}\right)_{p,\omega,t}.\label{ES18}
\end{align}

The Wigner transform of a convolution $A=B\circ C$ in time is not simply the product of the Wigner transforms of $B,C$. Instead one finds
\begin{align}
A_{\omega,t} =&\mathcal{WT}\{A\} = \mathcal{WT}\{B \circ C\}\label{ES19}\\
=&\mathcal{WT}\{B\} e^{\frac{i}{2}(\stackrel{\leftarrow}{\partial}_{t}\stackrel{\rightarrow}{\partial}_{\omega}-\stackrel{\leftarrow}{\partial}_{\omega}\stackrel{\rightarrow}{\partial}_{t})} \mathcal{WT}\{C\}\nonumber\\
=&B_{\omega,t} e^{\frac{i}{2}(\stackrel{\leftarrow}{\partial}_{t}\stackrel{\rightarrow}{\partial}_{\omega}-\stackrel{\leftarrow}{\partial}_{\omega}\stackrel{\rightarrow}{\partial}_{t})} C_{\omega,t}.\nonumber
\end{align}
Partial derivatives pointing to the left act on $B$, while those pointing on the right act on $C$. Thus approximating $A_{\omega,t}=B_{\omega,t}C_{\omega,t}$ by the product of Wigner transforms is ok, if the corrections to the zeroth order term in the exponential in Eq.~\eqref{ES19} are sufficiently small. 

The expansion of the kinetic equation in zeroth order is
\begin{align}
i\partial_{t} \tilde{F}_{p,\omega,t} \approx \Sigma^{R}_{p,\omega,t}\tilde{F}_{p,\omega,t}\Sigma_{z}-\Sigma_{z}\tilde{F}_{p,\omega,t}\Sigma^{A}_{p,\omega,t}+\Sigma^{K}_{p,\omega,t},\label{ES20}
\end{align}
where $\tilde{F}_{p,\omega,t}=\Sigma_{z}F_{p,\omega,t}\Sigma_{z}$ and Eq.~\eqref{ES20} contains only matrix multiplications. This is the Wigner approximation for the kinetic equation and it is valid in the case of negligible higher order corrections, i.e.
\begin{align}
\frac{|\partial_tF_{\alpha\beta,q,\omega,t}||\partial_{\omega}\Sigma^R_{\alpha\beta,q,\omega,t}|}{|F_{\alpha\beta,q,\omega,t}||\Sigma^R_{\alpha\beta,q,\omega,t}|} \ll 1.\label{ES21}
\end{align} This criterion is discussed in Eq.~\eqref{E27}.

\section{Diagrammatic Analysis of the Self-Energies}\label{AS3}
We give a brief summary of the diagrammatic approach for the computation of the self-energies and refer to Ref.~\cite{buchhold2015kinetic} for a detailed analysis.  We use the shorthand notation $P=(p,\omega,t)$ and $-P=(-p,-\omega,t)$.

\subsection{Selection rules for the diagrammatic representation of the self-energies}
The combination of resonant interactions, momentum conservation and energy conservation restricts the contributions to the self-energies to a certain subclass of diagrams \cite{buchhold2015kinetic}. A general one-loop diagram is depicted in Fig. \ref{FS1}. We identify each leg with a label for momentum, frequency and subsystem $\alpha,\beta \in \{1,2\}$. In addition, we explicitly label the direction of momentum and energy transfer with $r_{\alpha/\beta} \in \{ \pm 1\}$  for outgoing and incoming.

The resonance condition yields an additional constraint to this class of diagrams, which, after some algebra, requires
 $r_{\beta,k,\mu\phantom{\tilde{k}}}r_{\alpha,\tilde{k},\tilde{\mu}\phantom{\tilde{k}}}=r_{\beta,q,\nu\phantom{\tilde{k}}}r_{\alpha,\tilde{q},\tilde{\nu}\phantom{\tilde{k}}}=r_{\beta,p,\omega\phantom{\tilde{k}}}r_{\alpha,\tilde{p},\tilde{\omega}\phantom{\tilde{k}}}$. In addition to momentum and energy conservation at each vertex. This very helpful condition implies, that normal propagators contribute exclusively to the normal self-energy, while anomalous propagators contribute exclusively to the anomalous self-energy.

\begin{figure*}
\centering
\includegraphics[width=0.35\linewidth]{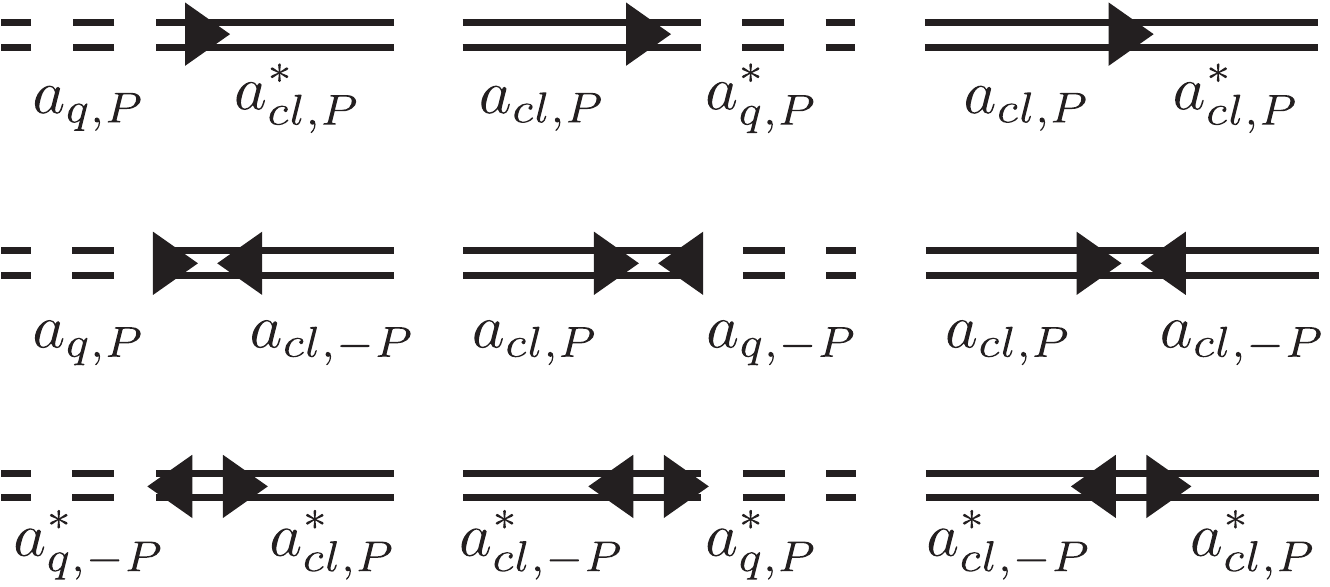}
\hspace{2cm}
\includegraphics[width=0.45\linewidth]{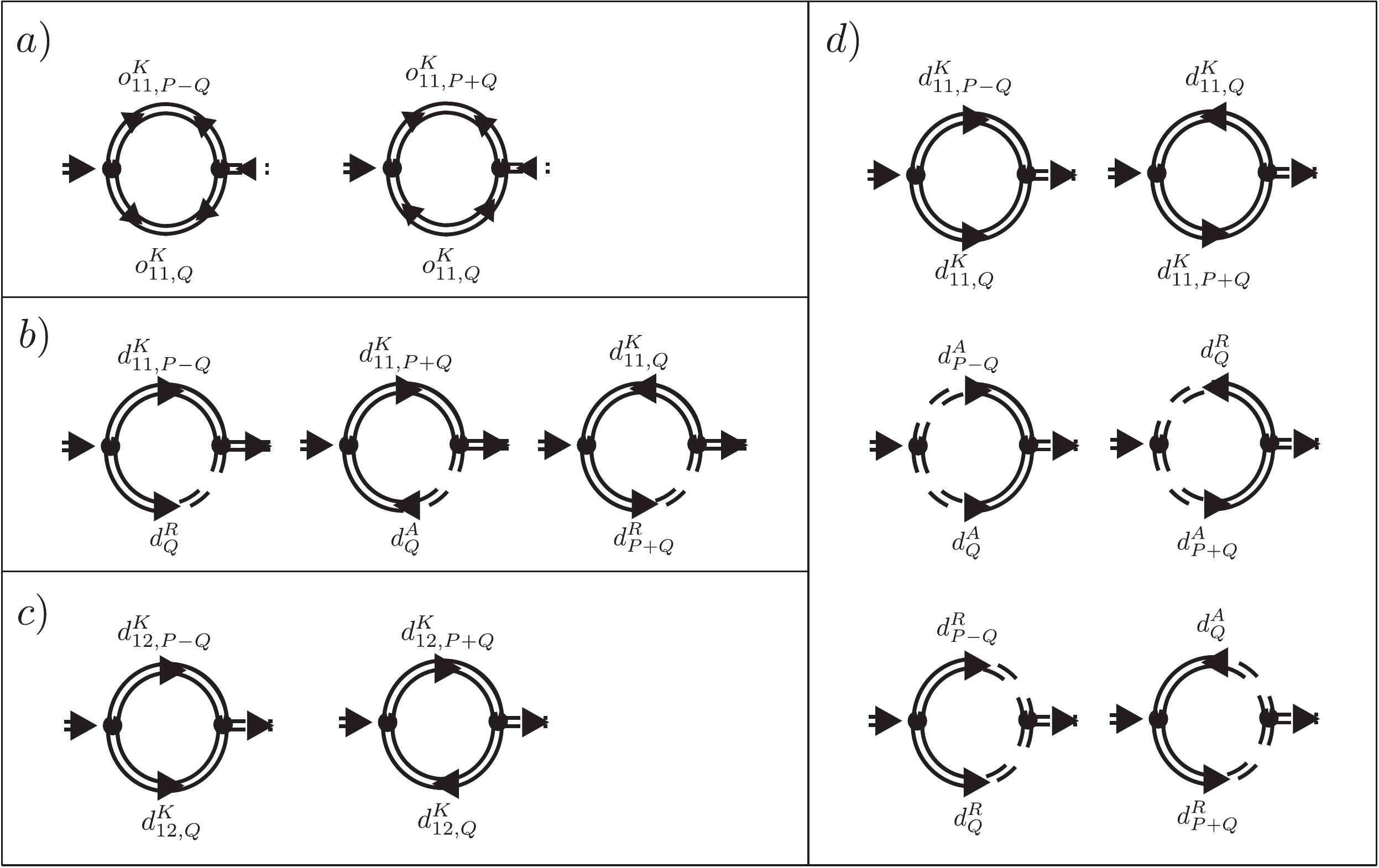}
\caption{Left column: Illustration of the diagrammatic lines for the propagators in Keldysh space. Dashed lines correspond to quantum fields and bold lines to classical fields. Arrows indicate the momentum and energy transfer direction, going from annihilation to creation operators. Right column: Diagrammatic representation of the self-energy diagrams. a) Anomalous intra-wire Keldysh self-energy, b) Retarded self-energy, c): Normal inter-wire  Keldysh self-energy, d): Normal intra-wire Keldysh self-energy.}
\label{FS3}
\end{figure*}
\newpage
\subsection{Self-energies}
The self-energies are determined via the canonical Dyson-Schwinger approach \cite{buchhold2015kinetic}. The corresponding diagrams are summarized in Fig.~\ref{FS3}. The analytic expressions for the self-energies contain the diagonal Green's functions $d^{R,A,K}$ and the off-diagonal Green's functions $o^K$ as defined in Eqs.~\eqref{ES4}, \eqref{ES5}, as well as the vertex $V_{p,q,p\pm q}$. The Dyson-Schwinger equations for the latter can be found in Ref.~\cite{buchhold2015kinetic}. Using their bare value (which corresponds to the Born approximation) yields $V(p,q,p\pm q)=\sqrt{|pq(p\pm q)|}$. The Dyson-Schwinger equations for the self-energies are (all retarded and advanced functions are diagonal in wire-index and Nambu space)
\begin{widetext}
\begin{align}
\sigma_{P}^{R}=\frac{i}{4\pi^2}\int_{q}\int_{\nu}&\Big(2V_{p,q,p- q}^{2}[d^{K}_{11,P-Q}d_{Q}^{R}]+2V_{p,q,p+q}^{2}[d^{K}_{11,P+Q}d_{Q}^{A}]+2V_{p,q,p+q}^{2}\ [d^{R}_{P+Q}d_{11,Q}^{K}]\Big),\\
\sigma^{K}_{11,P}=-\frac{1}{4\pi^2}\int_{q}\int_{\nu}&\Big(V_{p,q,p- q}^{2}\ [d_{P-Q}^{R}d_{Q}^{R}]+V_{p,q,p- q}^{2}[d_{P-Q}^{A}d_{Q}^{A}]+V_{p,q,p- q}^{2}[d^{K}_{11,P-Q}d^{K}_{11,Q}]\\
+&2V_{p,q,p+q}^{2}[d_{P+Q}^{R}d_{Q}^{A}]+2V_{p,q,p+q}^{2}[d_{P+Q}^{A}d_{Q}^{R}]+2V_{p,q,p+q}^{2}[d^{K}_{11,P+Q}d^{K}_{11,Q}]\Big),\nonumber\\
\sigma^{K}_{12,P}=-\frac{1}{4\pi^2}\int_{q}\int_{\nu}&\Big(V_{p,q,p- q}^{2}[d^{K}_{12,P-Q}d^{K}_{12,Q}]+2V_{p,q,p+q}^{2}[d^{K}_{12,P+Q}d^{K}_{12,Q}]\Big),\\
\Phi^{K}_{11,P}=-\frac{1}{4\pi^2}\int_{q}\int_{\nu}&\Big(V_{p,q,p- q}^{2}[o^{K}_{11,P-Q}o^{K}_{11,Q}]+2V_{p,q,p+q}^{2}[o^{K}_{11,P+Q}o^{K}_{11,Q}]\Big).
\end{align}
\end{widetext}

In the quasi-particle approximation, all the spectral weight is located on-shell. As a consequence, the spectral function $\mathcal{A}_{P}$, defined in equation \eqref{E28}, is sharply peaked at frequencies equal to the dispersion. Then retarded, advanced and Keldysh Green's function can be expressed via the spectral function and the phonon densities, as in Eqs~\eqref{E28}, \eqref{E29}. This yields
\begin{widetext}
\begin{align}
\sigma_{p,t}^{R}=\int_{q}\int_{\omega,\nu}&\Big(V_{p,q,p- q}^{2}[2n_{1,1,p-q,t}+2n_{1,1,q,t}+2]\mathcal{A}_{P-Q}\mathcal{A}_{Q}+2V_{p,q,p+q}^{2}[2n_{1,1,q,t}-2n_{1,1,p+q,t}]\mathcal{A}_{P+Q}\mathcal{A}_{Q}\Big),\\
\sigma^{K}_{11,p,t}=\int_{q}\int_{\omega,\nu}&\Big(V_{p,q,p- q}^{2}[(4n_{1,1,p-q,t}n_{1,1,q,t}+2n_{1,1,p-q,t}+2n_{1,1,q,t}+2]\mathcal{A}_{P-Q}\mathcal{A}_{Q}\\
+&2V_{p,q,p+q}^{2}[4n_{1,1,p+q,t}n_{1,1,q,t}+2n_{1,1,p+q,t}+2n_{1,1,q,t}]\mathcal{A}_{P+Q}\mathcal{A}_{Q}\Big),\nonumber\\
\sigma^{K}_{12,p,t}=\int_{q}\int_{\omega,\nu}&\Big(V_{p,q,p- q}^{2}[4n_{1,2,p-q,t}n_{1,2,q,t}]\mathcal{A}_{P-Q}\mathcal{A}_{Q}+2V_{p,q,p+q}^{2}[4n_{1,2,p+q,t}n_{1,2,q,t}]\mathcal{A}_{P+Q}\mathcal{A}_{Q}\Big),\\
\Phi^{K}_{11,p,t}=\int_{q}\int_{\omega,\nu}&\Big(V_{p,q,p- q}^{2}[4m_{1,1,p-q,t}m_{1,1,q,t}]\mathcal{A}_{P-Q}\mathcal{A}_{Q}+2V_{p,q,p+q}^{2}[4m_{1,1,p+q,t}m_{1,1,q,t}]\mathcal{A}_{P+Q}\mathcal{A}_{Q}\Big).
\end{align}\end{widetext}
This eliminates the frequency integrals from the expression and leads to Eqs.~\eqref{E33}-\eqref{E35}.

\end{document}